\documentclass[aps,pra,superscriptaddress,showpacs,reprint,nofootinbib]{revtex4-1}
\usepackage{graphicx}
\usepackage{mathrsfs}
\usepackage{amsmath}
\usepackage{subfigure}
\usepackage{bm}
\usepackage{verbatim}
\usepackage[colorlinks=true,letterpaper=true, pdfstartview=FitV,urlcolor=blue,linkcolor=orange,citecolor=cyan]{hyperref}
\usepackage{color}
\usepackage{xcolor}
\usepackage{hyperref}
\usepackage{amssymb}
\usepackage{bbm}
\usepackage{siunitx}
\usepackage{txfonts}

\newcommand{\bk}{\bm{k}}
\newcommand{\bK}{\bm{K}}

\usepackage{mathtools}
\DeclarePairedDelimiter\bra{\langle}{\rvert}
\DeclarePairedDelimiter\ket{\lvert}{\rangle}

\DeclarePairedDelimiterX\braket[2]{\langle}{\rangle}{#1 \delimsize\vert #2}
\DeclarePairedDelimiterX\inner[2]{\langle}{\rangle}{#1,#2}

\newcommand{\neiji}[2]{\langle #1|#2\rangle}

\newcommand{\br}{\bm{r}}

\usepackage{siunitx}

\newcommand{\TwoDMatrix}[4]{\begin{pmatrix} #1 & #2\\#3 & #4\end{pmatrix}}

\newcommand{\asi}{a_{\text{Si}}}
\newcommand{\EVS}{E_\text{VS}}

\newcommand{\tpsiL}{\tilde{\psi}_L}
\newcommand{\tpsiR}{\tilde{\psi}_R}
\newcommand{\tpsiLup}{\tilde{\psi}_{L\uparrow}}
\newcommand{\tpsiLdown}{\tilde{\psi}_{L\downarrow}}
\newcommand{\tpsiRup}{\tilde{\psi}_{R\uparrow}}
\newcommand{\tpsiRdown}{\tilde{\psi}_{R\downarrow}}
\newcommand{\tildet}{\tilde{t}_0}

\DeclareMathOperator\erfc{erfc}
\DeclareMathOperator\erf{erf}
\DeclareMathOperator\sech{sech}

\begin{document}
\title{
Transporting a single-spin qubit through a double quantum dot
}
\author{Xiao Li}
\email{lixiao@umd.edu}
\affiliation{Condensed Matter Theory Center and Joint Quantum Institute, University of Maryland, College Park, Maryland 20742, USA}
\author{Edwin Barnes}
\affiliation{Condensed Matter Theory Center and Joint Quantum Institute, University of Maryland, College Park, Maryland 20742, USA}
\affiliation{Department of Physics, Virginia Tech, Blacksburg, Virginia 24061, USA}
\author{Jason P. Kestner}
\affiliation{Department of Physics, University of Maryland Baltimore County, Baltimore, Maryland 21250, USA}
\author{S. Das Sarma}
\affiliation{Condensed Matter Theory Center and Joint Quantum Institute, University of Maryland, College Park, Maryland 20742, USA}

\date{\today}

\begin{abstract}
Coherent spatial transport or shuttling of a single electron spin through semiconductor nanostructures is an important ingredient in many spintronic and quantum computing applications.
In this work we analyze the possible errors in solid-state quantum computation due to leakage in transporting a single-spin qubit through a semiconductor double quantum dot. 
In particular, we consider three possible sources of leakage errors associated with such transport: finite ramping times, spin-dependent tunneling rates between quantum dots induced by finite spin-orbit couplings, and the presence of multiple valley states. In each case we present quantitative estimates of the leakage errors, and discuss how they can be minimized. 
Moreover, we show that in order to minimize leakage errors induced by spin-dependent tunnelings, it is necessary to apply pulses to perform certain carefully designed spin rotations. We further develop a formalism that allows one to systematically derive constraints on the pulse shapes and present a few examples to highlight the advantage of such an approach. 
\end{abstract}

\maketitle

\section{Introduction}

It has long been recognized that the ability to efficiently transport spins is crucial in designing spintronic devices or practical quantum computing in a physical system with short-range interactions~\cite{Ono2005}.  
Semiconductor spin qubits, including coupled donor spins~\cite{Kane1998,Vrijen2000electron} as well as electron spins localized in laterally defined quantum dots~\cite{Loss1998,Oosterkamp1998:QuantumDot,Burkard1999:QuantumDots,Hu2000hilbert,Hu2001spin,Hu2006:SpinQubit}, are prime examples of a candidate quantum information processor that typically has short-range spin-spin interactions, either exchange~\cite{Loss1998,Kane1998} or capacitive~\cite{Shulman2012}.  
Spin qubits have remarkably long coherence times compared to their gate times~\cite{Bluhm2011, Veldhorst2014}, and have been operated in one-dimensional arrays~\cite{Shulman2012,Braakman2013,Takakura2014}, and could potentially be operated in two-dimensional arrays~\cite{DiVincenzo2000,Hollenberg2006,cole2008spatial,Hill2015}. 
However, the ability to transport a single electron spin through semiconductor nanostructures while preserving its spin state has only been experimentally investigated recently~\cite{Mcneil2011,Baart2016}.

One way or another, spin shuttling schemes have always featured in theoretical considerations of scalable spin qubit architectures~\cite{Taylor2005,Hollenberg2006}. Especially in recent proposals based on the two-dimensional surface code~\cite{Fowler2012}, it becomes necessary to move spins either as probes~\cite{OGorman2014}, or as qubit activators~\cite{Hill2015}, or as qubits themselves~\cite{Pica2015}.  
Mathematically, the Hilbert space of the entire system is composed of many orthogonal two-dimensional subspaces with spatially localized support, and initializing a physical electron spin into a given dot/donor not only populates one of those subspaces, but labels it as the ``logical'' subspace (or, in a multi-qubit system, part of the logical subspace) where the information resides.  
The subspaces with support on unpopulated dots/donors are labeled as ``leakage'' subspaces, as stray populations leaking into this space comes at the expense of the logical subspace.  A deliberate and \emph{complete} physical transport of a spin from one dot/donor to another is simply a repartitioning of the total space into relabeled logical and leakage subspaces, bringing the qubit into a potentially advantageous spatial position for storage, manipulation, or readout. However, in practice, any transport process will be susceptible to leakage, i.e., the \emph{incomplete} transfer of population to the new subspace, in which case the portion remaining in the original position is no longer part of the logical subspace.  
Such leakage processes may further be complicated by the presence of spin-orbit coupling (SOC)~\cite{Bulaev2005:QuantumDot,Stano2005:QuantumDot,Danon2009,Schreiber2011,Stepanenko2012,Yang2013:SiliconCubit,Maisi2015:SpinQubit,Veldhorst2015SOC} and multiple valley states~\cite{Boykin2004:ValleySplitting,Culcer2009:SpinQubit,Saraiva2009:SpinQubit,Li2010:Qubits,Culcer2010:SiliconQubit,Friesen2010:valley-orbit,Culcer2010:Cubit,Saraiva2011:SpinQubit}. 
Obviously, the whole Hilbert space of the system is typically much larger than the ``logical subspace'' of interest, and unless very special care (involving adiabaticity and other constraints) is taken, there would invariably be undesirable leakage, adversely affecting the whole spin shuttling scheme.  The quantitative details of such leakage will depend on the details of the physical system as well as the transport process with many mechanisms present in real semiconductors potentially playing detrimental roles.  In spite of there being many proposals in the theoretical literature for spin shuttling in semiconductor nanostructures, there has not been any detailed theoretical study of the leakage problem, which, however, now takes on significance in view of the recent experiments on spin transport in quantum dots~\cite{Mcneil2011,Baart2016}. 

In this paper we present comprehensive calculations of this leakage when transporting a single-spin qubit through a double quantum dot using a minimal model that captures the essential physics. 
Our purpose is to identify and analyze common sources of leakage errors during such an operation, and discuss how they can be possibly minimized. 
In particular, this work focuses on leakage errors that arise due to finite ramping times, spin-dependent tunneling rates induced by finite spin-orbit couplings, as well as the presence of multiple valley states. 
We first take into account the finite bandwidth of gate voltage pulses, as indeed the nonzero ramping time has important ramifications here. We show that even in the absence of any coupling between spin and orbital degrees of freedom, leakage occurs due to imperfect initialization to a localized Fock-Darwin state~\cite{Burkard1999:QuantumDots}. 
However, such ramping induced leakage errors can be compensated during the transport by adjusting the tunneling time appropriately.  

We perform a similar analysis in the presence of spin-orbit couplings, in which case the spin-dependence of the tunneling rate precludes the same compensation procedure, especially when the spin encodes a quantum bit in an arbitrary superposition state. 
We show that in such cases it is necessary to apply carefully designed spin rotations during the transport process to help preserve the spin state. We developed a systematic approach to determine pulse shapes that can fulfill such requirements. In particular, our approach does not rely on the prior knowledge of the spin state being transported, which is often a huge benefit. 
Sometimes it is also desirable to perform certain spin rotations while transporting the spin. However, due to the spin dependent tunneling rates, it is again necessary to choose the pulse shapes carefully in order to minimize the errors of the spin rotations performed. We show that the approach we established before can also help us construct pulse shapes that minimize the errors. 
Developing physical error correction protocols by designing appropriate spin rotations in the presence of spin-orbit coupling is an important new result in this paper.

We also present a discussion of the effect of multiple valley states, which is common in silicon quantum dots. Silicon is an outstanding candidate for the operation of spin qubits because of its long spin coherence time (because of the relative absence of spinful nuclei in the environment) and the potential for scalability (because of the vast existing silicon technology)~\cite{Tyryshkin2006,Pla2012,Kawakami2014,Veldhorst2015}. In addition, the spin-orbit coupling is generally weak in silicon, so that the information stored in the spin is relatively immune to charge noise~\cite{Zwanenburg2013silicon}. Furthermore, the nuclear-spin background fluctuations can also be largely suppressed by isotope engineering~\cite{Itoh2014isotope}. 
All these advantages have driven a concerted research effort in silicon spin qubits~\cite{Maune2012,Kim2014}. 
Despite all these advantages, however, silicon has one serious disadvantage (compared, for example, with GaAs) for quantum computing applications since it has a conduction band valley degree of freedom which complicates initialization and gate operations unless the valley degeneracy is somehow lifted by an energy splitting larger than the operational temperature.  
In addition, one may wonder whether the presence of multiple valley states could  lead to additional leakage channels, and thereby further complicate spin-qubit operations through the population of undesirable  valley states. 
Our study shows that different valley eigenstates in a pristine~\footnote{The word ``pristine'' in this paper specifically refers to a system with zero spin-orbit couplings.} silicon quantum dot are orthogonal to each other, and thus will not become a source of leakage for the spin qubit transport and operations. 
However, recent experiments~\cite{Yang2013:SiliconCubit} show that certain mechanisms like spin-orbit couplings may induce nonzero couplings between valley eigenstates. In such scenarios leakage to the other valley eigenstates indeed becomes possible. However, these couplings are appreciable only near certain so-called ``leakage hot spots''~\cite{Yang2013:SiliconCubit}. We demonstrate that as long as one can avoid such ``hot spots'', the leakage to the other valley states will still be negligible. As a result, the presence of multiple valley states in silicon quantum dots will usually not cause additional leakage errors in spin qubit transport and operations, although one needs to keep in mind the possibility of valleys causing problems through leakage hot spots in the presence of spin-orbit coupling.

The structure of this paper is as follows. 
In Section~\ref{Section:GaAs} we discuss the leakage in transporting a single-spin qubit through a double quantum dot in GaAs, at first neglecting spin-orbit coupling effects completely. Because there are no spin-orbit couplings or additional valley states, the leakage here is solely due to finite ramping times. This relatively simple situation allows us to set up the general framework for leakage calculations. 
In Section~\ref{Section:SOC} we add spin-orbit couplings to the GaAs double quantum dot model, and present a comprehensive analysis of leakage errors induced by finite spin-orbit couplings. In particular, we describe in some details on how to systematically design appropriate pulse shapes to mitigate the leakage. 
In Section~\ref{Section:Silicon} we turn to the leakage in silicon double quantum dots due to the multiple valley states, and discuss how one can usually avoid the leakage near the so-called ``leakage hot spots''. 
We summarize our results and discuss open questions in Section~\ref{Section:Summary}.

\section{Leakage in a double quantum dot in GaAs \label{Section:GaAs}}
We begin our analysis with the transport of a single-spin qubit through a GaAs double quantum dot (DQD). Because there are no additional valley states in GaAs quantum dots and we neglect spin-orbit couplings in this section, the leakage is entirely due to the finite ramping time that causes an imperfect initialization of the spin qubit. This relatively simple case allows us to set up the basic framework to analyze the leakage problem, and pave the way for more general discussions later. 

\subsection{Model for a double quantum dot} 

\begin{figure}[!]
\includegraphics[scale=1]{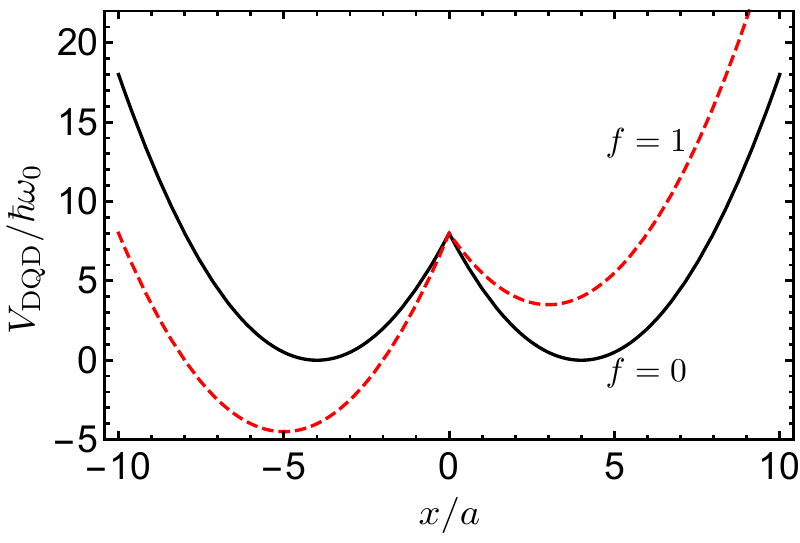}
\caption{\label{Fig:DQD} Illustration of the double quantum dot potential $V_\text{DQD}$ defined in Eq.~\eqref{Eq:DQD-Potential}. The dimensionless separation between the two dots is $\lambda = 4$, and $f$ is the quantum dot detuning parameter [see Eq.~\eqref{Eq:DetuningParmater}].} 
\end{figure}

We will first construct the model for a double quantum dot. If one neglects the finite confinement length in the $\hat{z}$-direction (which is justified since the typical $\hat{z}$-confinement is much tighter than the lateral lithographic confinement), the model for a double quantum dot can be cast in the following strictly two-dimensional form~\cite{Burkard1999:QuantumDots,Hu2006:SpinQubit}, 
\begin{align}
H_\text{DQD} = T + V_\text{DQD}(\br), \label{Eq:DQDHamiltonian}
\end{align}
where $T$ is the kinetic energy, and $V_\text{DQD}(\br)$ is the two-dimensional confinement potential of the double quantum dot, 
\begin{align}
V_\text{DQD}(\br) = \dfrac{m\omega_0^2}{2}\left[(|x|-X_0)^2+y^2\right]+eEx, \label{Eq:DQD-Potential}
\end{align}
where $E$ is an electric field along the interdot axis $\hat{x}$ that controls the detuning between the two quantum dots, $(\pm X_0, 0)$ is the location of the two dots in the $E=0$ limit, and $\hbar\omega_0$ is the strength of the harmonic confinement potential, which is usually a few meV. An illustration of $V_\text{DQD}$ is shown in Fig.~\ref{Fig:DQD}. 
Our results change little if the two-dimensional confinement is taken to be somewhat different from the parabolic confinement model. 

In the zero detuning limit ($E=0$), the confinement potential $V_\text{DQD}$ near each dot is harmonic at the center, and can be locally approximated by 
\begin{align}
V_{0}^{(\pm)}(\br) = \dfrac{m\omega_0^2}{2}\left[(x\pm X_0)^2+y^2\right]. 
\end{align}
As a result, in the large separation limit ($X_0\to\infty$) the single-particle orbitals near each quantum dot can be well approximated by the so-called Fock-Darwin (FD) states~\cite{Burkard1999:QuantumDots}. 
Moreover, because in a typical quantum dot the lowest FD level is separated from the excited ones by at least few meV~\cite{Burkard1999:QuantumDots}, the single-spin qubit can be well defined within the two spin states of the lowest FD level in each dot, given by 
\begin{align}
\psi_{L/R}(x,y) = \dfrac{1}{\sqrt{\pi}a}e^{-[(x\pm X_0)^2+y^2]/2a^2}, \label{Eq:FDStates}
\end{align}
where $a = \sqrt{\hbar/m\omega_0}$ is the FD radius. 
We want to emphasize that these two basis states generally have a nonzero overlap, 
\begin{align}
	l \equiv \neiji{\psi_{L}}{\psi_{R}} = e^{-\lambda^2}, \label{Eq:Overlap}
\end{align}
where $\lambda = X_0/a$ is the dimensionless separation between the two quantum dots. 
Note also that the above FD wave functions and their overlap will both be modified in the presence of an out-of-plane magnetic field (i.e., applied along the $z$ direction)~\cite{Li2010:Qubits}. However, in this work we  limit our discussions to the zero field case, because our results will not be qualitatively modified by the magnetic field~\footnote{We do take into account the effect of the finite magnetic field explicitly in the next section where we consider leakage errors in the presence of spin-orbit couplings because the time reversal breaking by the applied magnetic field introduces qualitative issues when combined with the inversion symmetry breaking associated with the spin-orbit coupling.}. 

A finite electric field $E$ applied along the interdot axis will create a detuning $\varepsilon_d = 2eEX_0$ between the two  dots, and shift the minima of the double well potential by an amount $fa$, with the dimensionless detuning parameter $f$ given by~\cite{Li2010:Qubits}
\begin{align}
	f = eEa/\hbar\omega_0. \label{Eq:DetuningParmater}
\end{align}
Such an effect is illustrated by the dashed line in Fig.~\ref{Fig:DQD}. 
The displaced FD wave function can be obtained by replacing $\pm X_0$ by $\pm X_0-fa$ in Eq.~\eqref{Eq:FDStates}. As a result, neither the interdot separation nor the overlap between the FD wave functions is affected by the applied electric field. 

It is often more convenient to work with basis states that are orthogonal with each other. Therefore we introduce the so-called Wannier (single-dot) states as follows~\cite{Burkard1999:QuantumDots},  
\begin{align}
 \ket{\tpsiL} = \dfrac{\ket{\psi_{L}}-g\ket{\psi_{R}}}{\sqrt{1-2gl+g^2}}, \quad
 \ket{\tpsiR} = \dfrac{\ket{\psi_{R}}-g\ket{\psi_{L}}}{\sqrt{1-2gl+g^2}}, \label{Eq:WannierStates}
\end{align}
which are orthogonal $\neiji{\tpsiL}{\tpsiR}=0$, with $g = l/(1+\sqrt{1-l^2})$. 
Inversely, the FD states are written in the Wannier basis as 
\begin{align}
	\ket{\psi_{L}} = \cos\dfrac{\theta}{2}\ket{\tpsiL} + \sin\dfrac{\theta}{2}\ket{\tpsiR}, \;
	\ket{\psi_{R}} = \sin\dfrac{\theta}{2}\ket{\tpsiL} + \cos\dfrac{\theta}{2}\ket{\tpsiR}, \notag
\end{align}
where $\theta = \arcsin l$. 

In this Wannier basis the double quantum dot Hamiltonian has the following form, 
\begin{align}
H_\text{Wannier} =
	\begin{pmatrix}
		\tilde{\varepsilon}_d/2 & \tildet \\ \tildet & -\tilde{\varepsilon}_d/2
	\end{pmatrix}, \label{Eq:WannierH}
\end{align}
where $\tilde{\varepsilon}_d \equiv \bra{\tpsiL} H_\text{DQD}\ket{\tpsiL} - \bra{\tpsiR} H_\text{DQD}\ket{\tpsiR}$ is the orbital detuning between the two quantum dots, while ${\tildet}\equiv \bra{\tpsiL} H_\text{DQD}\ket{\tpsiR}$ captures the tunneling between the two quantum dots. The explicit expressions for these two parameters are~\cite{Li2010:Qubits}
\allowdisplaybreaks[4]
\begin{align}
\tilde{\varepsilon}_d &= \dfrac{-\lambda^3\epsilon_0}{\sqrt{\pi(1-l^2)}}\biggl[e^{-(\lambda+f)^2}-e^{-(\lambda-f)^2}\notag\\&\quad +\sqrt{\pi}\left[(\lambda+f)\erf(\lambda+f)-(\lambda-f)\erf(\lambda-f)\right]\biggr],\notag\\
{\tildet}&= -\dfrac{\lambda^3 l\epsilon_0}{2(1-l^2)}\biggl[2\lambda+2f\erf(f)-(\lambda-f)\erf(\lambda-f)\notag\\&-(\lambda+f)\erf(\lambda+f)-\dfrac{e^{-(\lambda+f)^2}}{\sqrt{\pi}}\left(1+e^{4\lambda f}-2e^{\lambda(\lambda+2f)} \right) \biggr], 
\end{align}
where $\erf(x)$ is the error function, and $\epsilon_0 = \hbar^2/mX_0^2$ is a characteristic geometric energy scale determined solely by intrinsic quantum dot parameters. Using that $m = 0.065m_e$ in GaAs ($m_e$ is the bare electron mass) and choosing $X_0 = \SI{70}{nm}$, we find that $\epsilon_0 \simeq \SI{0.25}{meV}$. 
Note also that the tunneling parameter $\tildet$ is always negative. 
Finally, an FD state on the left (right) dot can be recovered as the ground state of $H_\text{Wannier}$ in the $f\to+\infty$ $(f\to-\infty)$ limit. 

\subsection{Leakage in transporting a single-spin qubit through a double quantum dot with no spin-orbit couplings}
We now set up the basic framework to study the leakage in the transport of a single-spin qubit through this double quantum dot. 
We consider that a single electron is first loaded from a reservoir onto the left quantum dot in the positively detuned limit $f\to +\infty$, which will guarantee that it starts in an exact FD state on the left dot.
Transporting the spin qubit then involves three successive stages: initialization (I), gate operation (II), and finalization (III). Specifically, in the initialization stage (I) the detuning is quickly reduced to zero within time $T_I$; 
in the gate operation stage (II), the electron tunnels into the right quantum dot within time $T_{II}$; 
and finally, in the finalization stage (III), the detuning between the dots is quickly increased to the negatively detuned limit $f\to -\infty$, in order to preserve the final state on the right dot. The ramping time in this stage usually equals $T_I$ in actual experiments. 

At the end of stage III, we can evaluate the overall error during the transport, defined as the error in obtaining the desired state on the right dot at the end of stage III, i.e., 
\begin{align}
	\eta_\text{overall} = 1-|\neiji{\psi_{f}}{\psi_\text{III}}|^2,
\end{align}
where $\psi_\text{III}$ is the state at the end of stage III, and $\psi_f$ is the desired final state.  

There are many possible sources of leakage errors. In this work we will focus on three typical ones. The first one arises because the ramping process cannot be instantaneous~\footnote{Strictly speaking, a truly instantaneous ramping involving zero ramping time will introduce errors associated with leakage into all the higher orbital excited states of the system since zero ramping couples excited states at all energies---we ignore these complications since they are not of any practical relevance for the real system.}. As a result, the state at the end of stage I may not be an exact FD state on the left dot to begin with. We call this type of error the \emph{ramping error} $\eta_\text{ramping}$, defined as the error in obtaining an FD state on the left dot at the end of stage I, i.e., 
\begin{align}
	\eta_\text{ramping} = 1-|\neiji{\psi_{L}}{\psi_{I}}|^2, 
\end{align}
where $\psi_I$ is the state at the end of stage I. Such an error can arise even without spin-orbit couplings or additional valley states, and will be the main focus of this section. 

In the presence of finite spin-orbit couplings, the spin and orbit dynamics will be entangled, and thus additional leakage errors could arise. For example, the initial spin state of the electron may not be preserved during the transport. Similarly, any spin operations performed during the transport may also be subject to errors. We will present a comprehensive analysis on this type of leakage errors in Section~\ref{Section:SOC}. 

The last type of leakage error we consider is due to the additional valley states that are present in, for example, silicon quantum dots. 
These valley states are usually quasidegenerate, which seems to be a source of leakage errors for spin qubit transport and operations. We will analyze this problem in some detail in Section~\ref{Section:Silicon}. 

In the present section, however, we will only consider leakage in GaAs quantum dots without spin-orbit couplings. In such a case the spin and orbital dynamics completely decouple, and thus any spin dynamics will not be affected by the transport. As a result, the only possible leakage comes from the ramping errors. We will therefore completely ignore the spin degrees of freedom in this section.

\subsubsection{Stage I and III: The ramping error}
\begin{figure}[!]
\includegraphics[scale=1]{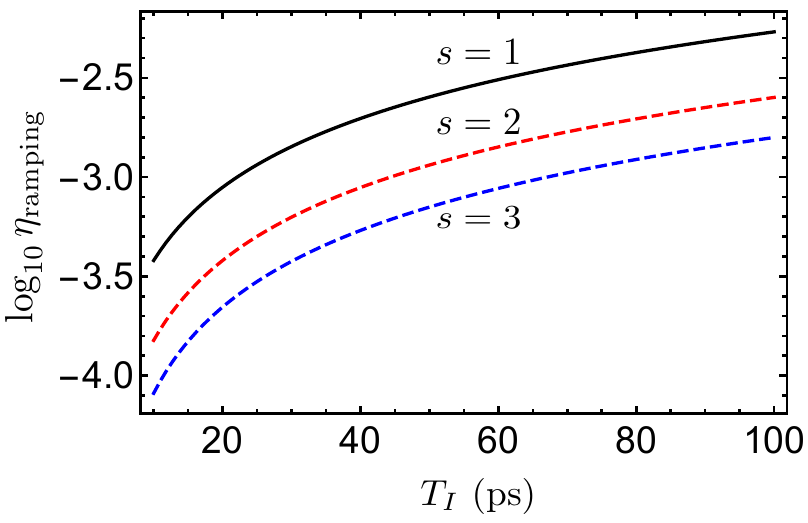}
\caption{\label{Fig:RampingError} Ramping errors as a function of ramping time $T_I$. In this figure the initial detuning is $f_0 = 4$, the dimensionless quantum dot separation is $\lambda = 2.0$, and the confinement potential has a strength of $\epsilon_0 = \SI{0.25}{meV}$. The three curves correspond to different time dependences of the detuning parameter, see Eq.~\eqref{Eq:RampingRate}. }
\end{figure}

We first study the ramping errors $\eta_\text{ramping}$ in the initialization stage (I). We assume that the detuning parameter $f(t)$ has the following time dependence, 
\begin{align}
	f(t) = f_0 (1-t^s/T_I^s), \label{Eq:RampingRate}
\end{align}
where $f_0$ is the initial detuning, $T_I$ is the ramping time, and $s$ allows for the possible nonlinear ramping rates. Our first observation is that a detuning parameter $f_0\geq 4$ is already sufficient to guarantee that the ground state $\psi_\text{GS}$ of the Wannier Hamiltonian~\eqref{Eq:WannierH} has an error less than $10^{-9}$ from an exact FD state on the left dot. 
We then numerically calculate the time evolution of $\psi_\text{GS}$ by solving the Schroedinger equation for its time evolution operator $U(t)$ as follows, 
\begin{align}
	i\hbar\dfrac{\partial U(t)}{\partial t} = H_\text{Wannier}(t) U(t). 
\end{align}
The ramping error then amounts to 
\begin{align}
	\eta_\text{ramping} = 1 - |\neiji{\psi_{L}}{\psi_{I}}|^2 = 1-|\bra{\psi_{L}}U(T_I)\ket{\psi_\text{GS}}|^2, 
\end{align}
which is plotted as a function of ramping time $T_I$ in Fig.~\ref{Fig:RampingError}. We find that if the ramping process is linear $(s=1)$ the ramping error is typically larger than $10^{-3}$. In contrast, for a nonlinear ramping with $s > 1$, the ramping error can be much smaller. One possible explanation for this could be that for nonlinear ramping processes, a larger portion of the dynamics occurs in the larger detuning limit, so that the leakage to the right dot is relatively suppressed. 

In the third stage we drive the double quantum dot into a negatively detuned limit $(f\to-\infty)$, so that the FD state on the right dot can be preserved. The following dependence of $f(t)$ will be adopted, 
\begin{align}
	f(t) = f_0 (t^s/T_I^s). 
\end{align}
Because this stage is the reverse of stage I, we will not elaborate on it further. 

\subsubsection{Stage II: the Rabi oscillation process}
\begin{figure}[!]
\includegraphics[scale=1]{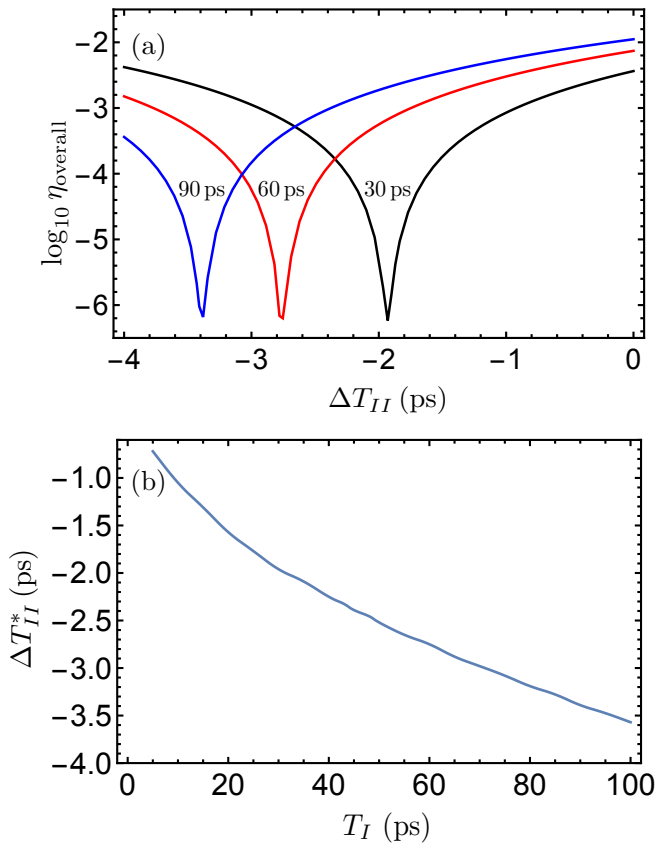}
\caption{\label{Fig:FinalError} (a) The overall error $\eta_\text{overall}$ as a function of the time adjustment $\Delta T_{II}$. The labels indicate the corresponding ramping time. (b) The optimal time adjustment $\Delta T_{II}^{\ast}$ as a function of the ramping time $T_I$. The other parameters in this plot are $f_0 = 4$, $\lambda = 2$, and $\epsilon_0 = \SI{0.25}{meV}$. Moreover, we used a linear ramping process [$s=1$ in Eq.~\eqref{Eq:RampingRate}]. }
\end{figure}

We now focus on the gate operation stage (II), where there is no detuning between the two quantum dots, so that the Hamiltonian in Eq.~\eqref{Eq:WannierH} reduces to $H_\text{Wannier} = \tildet^{(0)}\sigma_x$, with $\tildet^{(0)}$ being the tunneling parameter in the zero detuning limit, 
\begin{align}
\tildet^{(0)} = \dfrac{-\lambda^3 l\epsilon_0}{1-l^2}\left[\lambda\erfc(\lambda)+\dfrac{1-l}{\sqrt{\pi}}\right]. \label{Eq:tildet0}
\end{align}
As a result, the minimal time for an exact FD state on the left dot to tunnel to the right dot equals half the Rabi oscillation period, i.e., $T_{II}^{(0)} = T_\text{Rabi}/2 = {\pi\hbar}/{2|\tildet^{(0)}|}$, 
where the superscript $(0)$ indicates that this is the ideal operation time if the Rabi oscillation starts from an exact FD state on the left dot. 
For $\lambda = 2.0$ and $\epsilon_0 = \SI{0.25}{meV}$, we estimate that $\tildet^{(0)}\simeq \SI{-0.02}{meV}$, and thus $T_{II}^{(0)}\simeq\SI{51.70}{ps}$. 

However, because of the ramping errors, the state at the beginning of the Rabi oscillation will not be an exact FD state on the left dot. One may wonder how this ramping error affects the fidelity of obtaining an FD state on the right dot at the end of the operation, and how such errors can be corrected during the operations. 
Our calculations show that the overall error $\eta_\text{overall}$ can be much reduced if we slightly reduce the operation time $T_{II}$ in the second stage. In Fig.~\ref{Fig:FinalError}(a) we plot the overall error $\eta_\text{overall}$ as a function of the time adjustment $\Delta T_{II}$, defined as $\Delta T_{II} = T_{II} - T_{II}^{(0)}$,
where $T_{II}$ is the actual operation time in the second stage. We can see that if the operation time $T_{II}$ is reduced by an optimal amount $\Delta T_{II}^{\ast}$, the overall error can be minimized to be below $10^{-6}$. 
Figure~\ref{Fig:FinalError}(b) further demonstrates that 
the optimal operation time $T_{II}$ is shorter for larger ramping times $T_{I}$.
We also note that these results were obtained using a linear ramping process [$s=1$ in Eq.~\eqref{Eq:RampingRate}] in both stage I and III. If a nonlinear ramping process is carried out, the overall error can be further reduced. 

What we learn from the above analysis is that due to the ramping errors, the optimal operation time in the second stage should be reduced from half the Rabi oscillation period, i.e., $T_{II} = T_{II}^{(0)}+\Delta T_{II}^{\ast}$. One intuitive way to understand this result is the following. The Rabi oscillation from the left dot to the right dot is like flipping a ``spin-up'' state by a $\pi$-pulse. On a Bloch sphere this process is like to move the initial state from the north pole to the south pole. 
The initial ramping error causes the Rabi oscillation to start from somewhere away from the north pole, and thus it should take less time to arrive at the south pole. 




\subsection{Leakage in a DQD with smaller tunneling rates}

\begin{figure}[!]
\includegraphics[scale=1]{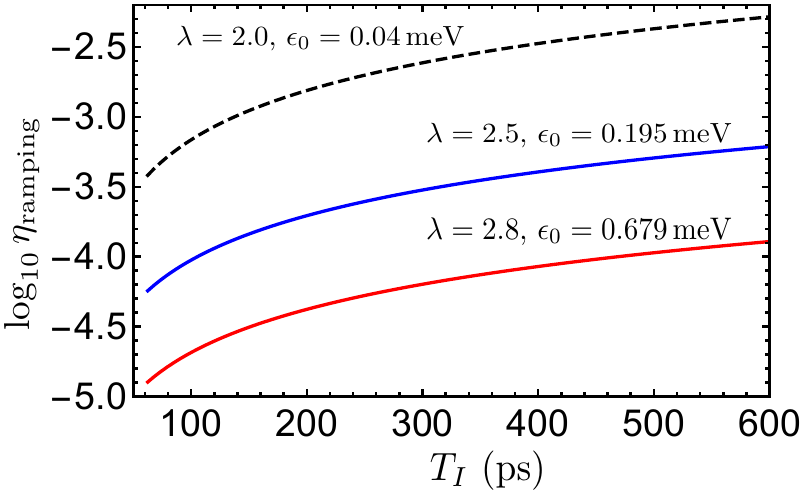}
\caption{\label{Fig:RampingErrorSmall} Ramping errors as a function of ramping time $T_I$. The tunneling rate at zero QD detuning is $\tildet^{(0)}=\SI{3.31}{\micro eV}$, and $f_0 = 4$. 
The three curves correspond to different ways to model the same $\tildet^{(0)}$. }
\end{figure}

We have been considering leakage in DQDs with relatively large tunneling rates, with $|\tildet^{(0)}|\simeq\SI{0.02}{meV}$ in the limit of zero orbital detuning. 
Our goal is to demonstrate that even relatively large leakage errors, i.e, the worst case scenario from the leakage perspective, can still be corrected. 
Nonetheless, it seems appropriate to present some quantitative  estimates using small tunneling rates closer to actual experimental values~\cite{Oosterkamp1998:QuantumDot,Schreiber2011,Maisi2015:SpinQubit,Baart2016}. 
In Fig.~\ref{Fig:RampingErrorSmall} we plot the ramping errors in a DQD with $|\tildet^{(0)}|=\SI{3.31}{\micro eV}$, which is close to the tunneling rate of $\SI{0.8}{GHz}$ measured in a recent experiment~\cite{Baart2016}. 
We find that a smaller tunneling rate generally allows for a longer ramping time in stage I while still keeping the ramping error low. 
The smaller tunneling rate also provides a longer gate operation time in stage II: with $|\tildet^{(0)}| = \SI{3.31}{\micro eV}$ we estimate that $T_{II}^{(0)} \simeq \SI{312.36}{ps}$. 
Thus, realistic tunneling rates allow for longer ramping times, which are still orders of magnitude longer than the qubit decoherence time ($>\SI{10}{\micro s}$ even in GaAs quantum dots).

There is an important caveat in modeling a given tunneling rate $\tildet^{(0)}$, though. As shown in Eq.~\eqref{Eq:tildet0} $\tildet^{(0)}$ depends on both $\lambda$ and $\epsilon_0$. Therefore in order to obtain a smaller $\tildet^{(0)}$ than what we use in the previous discussions, we can either fix $\lambda$ and just decrease $\epsilon_0$, or use a larger $\lambda$. 
The first case can be accommodated with a simple rescaling of the time variable in the Schroedinger equation. 
In particular, the dashed curve in Fig.~\ref{Fig:RampingErrorSmall} can be obtained from the black curve in Fig.~\ref{Fig:RampingError} by multiplying its $T_I$ variable with a factor of $\SI{0.25}{meV}/\SI{0.04}{meV}\simeq 6.25$. 
In the second case, however, by using a larger $\lambda$, the actual overlap between two quantum dots is reduced. In this case, the ramping error is even smaller, as shown by the two solid curves in Fig.~\ref{Fig:RampingErrorSmall}. 

Therefore we conclude that leakage errors in current experiments are relatively small due to the small tunneling rates between the quantum dots (at least as long as spin-orbit coupling effects are negligible). 
However, we will continue to use larger tunneling parameters in our analysis in order to better ascertain the quantitative extent to which leakage errors can be eliminated in quantum dot spin transport even in the worst case scenario. 


\section{Leakage in a double quantum dot with finite spin-orbit couplings \label{Section:SOC}}
The previous section focused on the leakage due to finite ramping times. In this section we turn to the second source of leakage errors, i.e., the presence of finite spin-orbit couplings. 
The entanglement of spin and orbital dynamics poses a substantial challenge in preserving the spin state during the transport process. 
We will show that except for some special cases, it is necessary to apply a carefully designed $2\pi$-pulse during the transport in order to help preserve the spin state. Fortunately, we find that there exists a systematic way to find $2\pi$-pulses that can fulfill such requirements. More importantly, such pulses can be designed without prior knowledge of the spin state being transported, which is highly desirable for quantum computation or quantum information processing purposes. 
We also consider the leakage errors when trying to perform certain spin rotations using an applied pulse while transporting the spin qubit. As expected, spin-orbit couplings often prevent such operations from being carried out exactly, and it is necessary to optimize the applied pulse in order to reduce errors. 
We show that the approach we develop can also provide crucial insights into how the pulse should be optimized, and we present some pertinent examples.  

The general form of spin-orbit couplings in a semiconductor quantum dot is given by~\cite{Bulaev2005:QuantumDot,Stano2005:QuantumDot,Yang2013:SiliconCubit}
\begin{align}
	H_{\text{SO}} = \beta_{D}(-s_x P_x + s_y P_y) + \alpha_{R} (s_x P_y - s_y P_x), \label{Eq:Spin-OrbitHamiltonian}
\end{align}
where the two terms correspond to the so-called Dresselhasus and Rashba contribution, respectively. The former appears when the inversion symmetry is broken in the bulk, while the latter appears in the presence of asymmetric confining potentials, explicitly breaking the spatial inversion symmetry. 
The effects of spin-orbit couplings on double quantum dots have been studied extensively in the literature~\cite{Stano2005:QuantumDot,Danon2009,Schreiber2011,Stepanenko2012}. 
In particular, many studies have demonstrated that the interdot tunneling $\tildet^{(0)}$ can be modified by the spin-orbit coupling, which will result in a quantitative modification of the leakage in the single-spin qubit operation we described previously. 

More interestingly, however, a finite magnetic field breaks the time-reversal symmetry, and allows for unequal tunneling rates between the two spin species. 
Quantitative estimates show that such a difference can be as large as $0.5\%$~\cite{Stano2005:QuantumDot}. 
Motivated by these considerations, we construct the following phenomenological model to capture such spin-dependent  tunneling rates induced by spin-orbit couplings, 
\begin{align}
H(t) = 
\begin{pmatrix}
 \dfrac{\tilde{\varepsilon}_d+E_z}{2} & \Omega(t)e^{-i\omega t}  & \tilde{t}_0(1+\zeta)  & 0 \\
\Omega(t)e^{+i\omega t}  & \dfrac{\tilde{\varepsilon}_d-E_z}{2} & 0 & \tilde{t}_0(1-\zeta)  \\
 \tilde{t}_0(1+\zeta)  & 0 & \dfrac{-\tilde{\varepsilon}_d+E_z}{2} & \Omega(t)e^{-i\omega t}  \\
 0 & \tilde{t}_0(1-\zeta)  & \Omega(t)e^{+i\omega t}  & \dfrac{-\tilde{\varepsilon}_d-E_z}{2} \end{pmatrix}, \label{Eq:HDriving}
\end{align}
which is written in the basis of $\{\tpsiLup, \tpsiLdown, \tpsiRup, \tpsiRdown\}$. Here $E_z$ is the Zeeman splitting, 
$\Omega(t)$ and $\omega$ are the magnetic field pulse envelope and frequency, respectively, while the phenomenological parameter $\zeta$ accounts for the difference in the two spin tunneling rates. 
Without any loss of generality we will assume $\zeta>0$ throughout our discussions. 
We can introduce a unitary transformation to remove the oscillatory time dependence in Eq.~\eqref{Eq:HDriving}, and obtain 
\begin{align}
H_\text{eff}(t) = 
\begin{pmatrix}
 \dfrac{\tilde{\varepsilon}_d+\delta}{2} & \Omega(t)  & \tilde{t}_0(1+\zeta)  & 0 \\
\Omega(t)  & \dfrac{\tilde{\varepsilon}_d-\delta}{2} & 0 & \tilde{t}_0(1-\zeta)  \\
 \tilde{t}_0(1+\zeta)  & 0 & \dfrac{-\tilde{\varepsilon}_d+\delta}{2} & \Omega(t)  \\
 0 & \tilde{t}_0(1-\zeta)  & \Omega(t)  & \dfrac{-\tilde{\varepsilon}_d-\delta}{2} \end{pmatrix}. \label{Eq:WannierFourLevel}
\end{align}
where $\delta = E_z-\omega$ is the detuning of the driving field $\Omega(t)$. 

\subsection{Leakage in transporting a spin-up or spin-down state \label{Section:SpinUp}}
We begin our analysis with the simple situation of transporting a spin-up or spin-down FD state. We will see that this special case avoids many subtle issues while still revealing the essence of the problem, thus giving us insight and intuition into the physics of the problem. We will turn to more general initial spin states at a later stage. 

\subsubsection{Stage I: The ramping error}
\begin{figure}[!]
\includegraphics[scale=1]{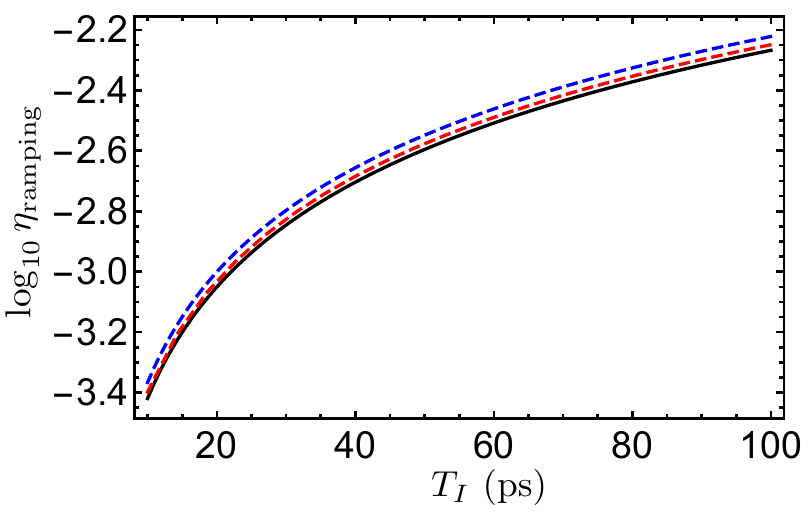}
\caption{\label{Fig:SOCRampingError} Ramping errors in initialing a spin-up FD state on the left dot. 
The black solid line corresponds to the zero SOC case (i.e., the same as the black line in Fig.~\ref{Fig:RampingError}), while the red and blue lines correspond to $\zeta = 0.02$ and $0.05$, respectively. The other parameters in this plot are $f_0 = 4$, $\lambda = 2$, and $\epsilon_0=\SI{0.25}{meV}$. }
\end{figure}

We first calculate the ramping errors when transporting a spin-up (or spin-down) electron in this model. Because no driving field is applied in the ramping process, the dynamics of spin-up and spin-down electrons are completely decoupled. 
As a result, starting with a spin-up or spin-down state has the important simplification that the spin state will not be affected by the ramping process.
We thus expect that the ramping error in this case only comes from the orbital part of the wave function, and it is larger (smaller) than that in a pristine system, because the effective tunneling rate between the two quantum dots is larger (smaller). 
The numerical result for ramping a spin-up FD state is given in Fig.~\ref{Fig:SOCRampingError}, which indeed shows that the corresponding ramping error increases with $\zeta$. 

\subsubsection{The overall error}

Next we discuss the overall error of transporting a spin-up FD state from the left dot to the right dot. 
We first consider a simple shuttling of the spin qubit, in which no driving field is applied in stage II to perform spin rotations. 
Consequently, the dynamics of spin-up and spin-down electrons is again decoupled, and thus this situation is almost the same as the one without SOC we discussed in the previous section: by properly adjusting the gate operation time $T_{II}$, one can perform the entire operation exactly. The only caveat is the definition of the time adjustment $\Delta T_{II}^{\ast}$. Here we choose to measure it with respect to $T_{II}$ in a system without SOC; as a result we expect that $\Delta T_{II}^{\ast}$ for a spin-up electron decreases with increasing $\zeta$, because the corresponding tunneling rate is increased. The numerical results in Fig.~\ref{Fig:SOCFinalError}(a) corroborate our expectations. 

A more interesting situation arises when a $\pi$-pulse is applied in stage II to flip the electron spin, in which case the error should be measured against a spin-down FD state on the right dot.  
In a system with zero SOC such an operation is still exact, because the spin rotations will not be affected by orbital dynamics. However, spin-orbit couplings will generally induce errors in such a combined operation, making it difficult to perform the desired operation exactly. 
In what follows we would like to quantify such errors and investigate if one can reduce them by optimizing pulse shapes. 

Our numerical results are shown in Fig.~\ref{Fig:SOCFinalError}(b). 
We first look at the three solid lines, which corresponds to using the following square $\pi$-pulse to flip the spin, 
\begin{align}
 	\Omega_\text{sq}(\tau) = \dfrac{\pi\hbar}{2T_{II}} \equiv \Omega_0. \label{Eq:SquarePulse}
 \end{align} 
We observe two prominent features in these curves. 
First, the optimal time adjustment $\Delta T_{II}^{\ast}$ in this case does not vary much with $\zeta$, in contrast to the case without a $\pi$-pulse [cf. Fig.~\ref{Fig:SOCFinalError}(a)]. 
We can understand this effect as follows: while a spin-up electron accumulates a larger ramping error in stage I than it does in the zero SOC case, in stage III it becomes a spin-down electron, and thus accumulates less ramping errors. As a result, the overall ramping error will be almost the same as that in the zero SOC case, leading to an almost identical time adjustment $\Delta T_{II}^\ast$. 
The second observation is that the leakage minimum generally increases with $\zeta$. In fact, as we demonstrate in the following, when using the square pulse in Eq.~\eqref{Eq:SquarePulse}, the leading-order leakage minimum is proportional to $\zeta^2$. 
However, it is difficult to gain additional insights into such leakage errors solely from the numerical results; in addition, they cannot provide explicit guidance on how to design pulse shapes that can minimize the leakage errors either. 

In the following we will turn to some analytical methods. In particular, we will use time-dependent perturbation theory to construct a formalism that helps us analyze the leakage errors. 
Such an approach is powerful in its broad generality, because it not only helps us understand why the leakage errors arise, but also reveals the explicit constraints that the applied $\pi$-pulse must satisfy in order to minimize the leakage errors. 
The two dashed lines in Fig.~\ref{Fig:SOCFinalError}(b) plot the leakage under an optimized $\pi$-pulse in Eq.~\eqref{Eq:SechPulse}, and we can see that the leakage error of this operation is indeed largely suppressed.

\subsubsection{Stage II: Leakage errors in gate operations}

\begin{figure}[!]
\includegraphics[scale=1]{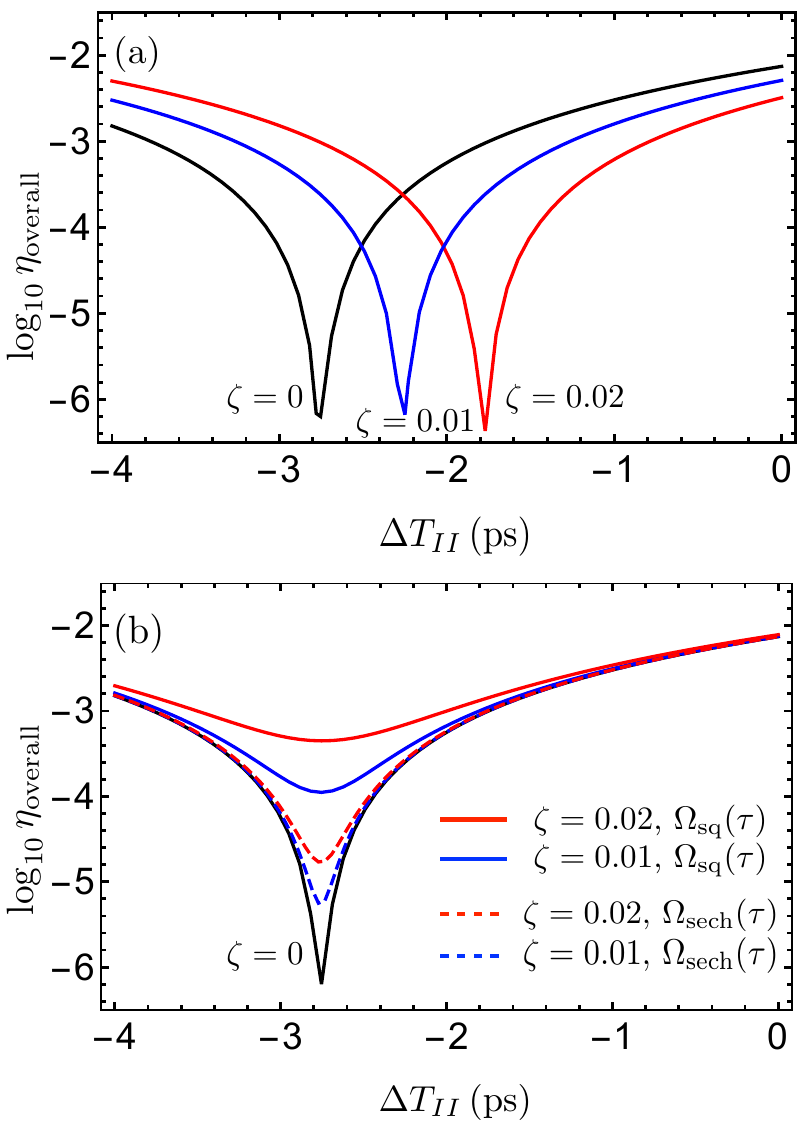}
\caption{\label{Fig:SOCFinalError} The overall error of transporting a spin-up FD state in the presence of finite spin-orbit couplings. 
(a) and (b) corresponds to the overall error without and with a $\pi$-pulse applied in stage II, respectively. The two pulse shapes considered in (b) are defined in Eq.~\eqref{Eq:SquarePulse} and~\eqref{Eq:SechPulse}, and they are both resonant with the Zeeman splitting. 
The parameter $\zeta$ captures the difference between the tunneling rates of spin-up and spin-down electrons [see Eq.~\eqref{Eq:WannierFourLevel}]. 
The time adjustment $\Delta T_{II}$ is measured with respect to the operation time $T_{II}^{(0)}$ in a pristine system $(\zeta=0)$. 
The other parameters are $T_I=\SI{60}{ps}$, $f_0=4$, $\lambda=2$, and $\epsilon_0 = \SI{0.25}{meV}$. }
\end{figure}

We begin our discussions by noting that the gate operation stage is carried out when the detuning between the two quantum dots is zero ($\tilde{\varepsilon}_d = 0$). Moreover, we will assume for the moment that the applied pulse is resonant with the Zeeman splitting ($\delta=0$), leaving the more general situation to the end of this section. 
Under such assumptions the Hamiltonian in Eq.~\eqref{Eq:WannierFourLevel} can be written as $H_\text{eff}(\tau) = H_0(\tau) + V$~\footnote{In this section we will use $\tau$ to represent the time variable, in order to avoid confusions with the tunneling rate $\tildet$.}, with
\begin{align}
	H_0(\tau) =\tildet \sigma_x\otimes s_0 + \Omega(\tau) \sigma_0\otimes s_x, \; V = \zeta\tildet \sigma_x \otimes s_z, \label{Eq:Hamiltonian-1}
\end{align}
where we adopt the convention that Pauli matrices $\sigma_{i}$ act on the orbital degrees of freedom, while $s_{i}$ act on the spin degrees of freedom. 
In the $\zeta=0$ limit, the time-evolution is only governed by $H_0(\tau)$, which can be solved exactly. Specifically, because the two terms in $H_0(\tau)$ commute with each other, the time-evolution operator for $H_0(\tau)$ can be written explicitly as 
\begin{align}
	U_0(\tau) = \exp\left[-i\tildet \tau\sigma_x\otimes s_0\right]\cdot\exp\left[-i\Phi(\tau)\sigma_0\otimes s_x \right], \label{Eq:Us1}
\end{align}
where $\Phi(\tau) = \int_0^{\tau}dt' \Omega(t')$ is the total area of the pulse. For example, a $\pi$-pulse will satisfy $\Phi(T_{II}) = \pi/2$, where again $T_{II}$ is the duration of stage II. 

The time evolution under $H_\text{eff}(\tau)$ with a nonzero $\zeta$ cannot be solved in a closed form. However, because $\zeta$ is generally small~\cite{Stano2005:QuantumDot} we can treat the $V$ term as a perturbation to $H_0(\tau)$. Such a problem can be handled most conveniently in the interaction picture~\cite{bruus2004many}. For clarity we will adopt the notation that operators and states carrying a hat symbol are in the interaction picture, while those without are in the usual Schroedinger picture. 
The operators and states in these two pictures are connected by $U_0(\tau)$ as follows, 
\begin{align}
	\hat{O}(\tau) = U_0^\dagger(\tau)O(\tau) U_0(\tau), \quad \ket{\hat{\psi}(\tau)} = U_0^{\dagger}(\tau)\ket{\psi(\tau)}. 
\end{align}
Moreover, by definition we have that $\hat{U}_0(\tau) = U_0(\tau)$, and that $\ket{\hat{\psi}(\tau=0)} = \ket{{\psi}(\tau=0)}$. Therefore we will not carry a hat symbol for $U_0(\tau)$. 

We can now write down the perturbation $V$ in the interaction picture as follows, 
\begin{align}
\hat{V}(\tau) = U_0^{\dagger}(\tau)V U_0(\tau)
= \zeta\tildet\sigma_x\otimes \left[\cos[2\Phi(\tau)]s_z + \sin[2\Phi(\tau)]s_y\right],  \notag
\end{align}
and the time-evolution operator in the interaction picture $\hat{U}(\tau)$ is only governed by $\hat{V}(\tau)$, i.e., 
\begin{align}
i\partial_\tau\hat{U}(\tau) = \hat{V}(\tau)\hat{U}(\tau). 
\end{align}
We can then solve $\hat{U}(\tau)$ up to second order in $\zeta$ as $\hat{U}(\tau) \simeq 1+\hat{U}^{(1)}(\tau) + \hat{U}^{(2)}(\tau)$, where
\begin{align}
\hat{U}^{(1)}(\tau) &= \dfrac{1}{i}\int_0^{\tau}d\tau_1\hat{V}(\tau_1), \notag\\
\hat{U}^{(2)}(\tau) &= \dfrac{1}{i^2}\int_0^{\tau}d\tau_1\hat{V}(\tau_1)\int_0^{\tau_1}d\tau_2\hat{V}(\tau_2). 
\end{align}
Finally, we can return to the Schroedinger picture and evaluate the overlap between any time-evolved initial state $\ket{\psi_i(\tau)}$ and the final state $\ket{\psi_{f}}$ as follows, 
\begin{align}
	\xi(\tau)\equiv\neiji{\psi_f}{\psi_i(\tau)} = \bra{\psi_f}U_0(\tau)\hat{U}(\tau)\ket{\psi_i(0)}. 
\end{align}

We will now use this formalism to analyze the leakage when flipping a spin-up state with a $\pi$-pulse while transporting it from the left dot to the right dot. We will demonstrate that by choosing an appropriate shape for the $\pi$-pulse, leakage errors up to second order in $\zeta$ can be eliminated. 

We first explore the condition to eliminate the first-order leakage error, which arises from $\hat{U}^{(1)}(\tau)$, 
\begin{align}
\hat{U}^{(1)}(\tau) = -i&\zeta\tildet\sigma_x\otimes \\&\biggl[s_z \int_0^{\tau}d\tau_1 \cos[2\Phi(\tau_1)]+ s_y \int_0^{\tau}d\tau_1 \sin[2\Phi(\tau_1)]\biggr]. \notag
\end{align}
Note that for a $\pi$-pulse, the time evolution operator $U_0(\tau)$ satisfies $U_0(\tau=T_{II})=-\sigma_x\otimes s_x$. Consequently, the contribution to the wave function overlap $\xi(\tau)$ from $\hat{U}^{(1)}(\tau)$ is 
\begin{align}
	\xi^{(1)}(T_{II}) &= \bra{\psi_R}\otimes\bra{\downarrow} \left[U_0(T_{II})\hat{U}^{(1)}(T_{II})\right]\ket{\uparrow}\otimes\ket{\psi_L} \notag\\
	&\propto \bra{\downarrow} s_x s_z \ket{\uparrow}\cdot \int_0^{T_{II}}d\tau_1 \cos[2\Phi(\tau_1)], 
\end{align}
where we have identified that the contribution from the other term in $\hat{U}^{(1)}(\tau)$ vanishes because $\bra{\downarrow}s_x s_y\ket{\uparrow}\equiv0$. As a result, in order to eliminate the first-order leakage error, we need a pulse that satisfies 
\begin{align}
	\int_0^{T_{II}}d\tau \cos[2\Phi(\tau)] = 0. \label{Eq:PulseShapeI}
\end{align}
Many $\pi$-pulses satisfy this constraint, including the square pulse in Eq.~\eqref{Eq:SquarePulse} and $\Omega_\text{sech}(t)$ we introduce in Eq.~\eqref{Eq:SechPulse} below. In fact, all $\pi$-pulses symmetric with respect to half of its period $T_{II}/2$ (or ``symmetric pulses'' for short), 
\begin{align}
\Omega(T_{II}/2-\tau) = \Omega(T_{II}/2+\tau), \quad (0<\tau<T_{II}/2), 
\end{align}
automatically satisfy this constraint. This conclusion can be proved by noting that all symmetric pulses satisfy $\Phi(T_{II}/2-t) = \pi/2-\Phi(T_{II}/2+t)$, and thus
\begin{align}
\int_0^{T_{II}/2}d\tau_1 \cos[2\Phi(\tau_1)] = -\int_{T_{II}/2}^{T_{II}}d\tau_1 \cos[2\Phi(\tau_1)]. 
\end{align}
One can also see from the proof that this conclusion holds for $\pi$-pulses only, while more general situations are considered below. 
In contrast, the following $\pi$-pulse is an example that cannot eliminate even the first-order leakage error, 
\begin{align}
	\Omega(\tau) = \dfrac{6\pi\hbar}{T_{II}^4}\tau^2(T_{II}-\tau). 
\end{align}

Next we study the condition to minimize the second-order leakage error, which arises from $\hat{U}^{(2)}(\tau)$, 
\begin{align}
\hat{U}^{(2)}(\tau)=-{\zeta^2\tildet^2} \int_0^{\tau}d\tau_1 &\int_0^{\tau_1}d\tau_2\biggl\{\cos[2\Phi(\tau_1)-2\Phi(\tau_2)]\notag\\&+is_x \sin[2\Phi(\tau_1)-2\Phi(\tau_2)] \biggr\}. 
\end{align}
Our first observation is that the contribution to $\xi^{(2)}(\tau)$ from the second term in $\hat{U}^{(2)}(\tau)$ is proportional to $\bra{\uparrow}s_xs_x\ket{\downarrow}$ and thus vanishes. 
As a result, the correction to the wave function overlap is given by 
\begin{align}
\xi^{(2)}(\tau) &= -\dfrac{\zeta^2\tildet^2}{2}\int_0^{\tau}d\tau_1 \int_0^{\tau}d\tau_2\cos[2\Phi(\tau_1)-2\Phi(\tau_2)], 
\end{align}
which then gives rise to the following leakage error 
\begin{align}
\eta^{(2)} = 1-|1+\xi^{(2)}(T_{II})|^2 \simeq 2|\xi^{(2)}(T_{II})|. 
\end{align}
It is also worth noting that the orbital part of the wave function will not carry additional leakage errors once the gate operation time is optimized. 
Therefore in order to minimize the leakage error at this order the pulse shape should satisfy 
\begin{align}
\int_0^{T_{II}}d\tau_1 \int_0^{T_{II}}d\tau_2 \cos[2\Phi(\tau_1)-2\Phi(\tau_2)] = 0. 
\end{align}
For symmetric $\pi$-pulses, which automatically satisfy the first condition in Eq.~\eqref{Eq:PulseShapeI}, the above condition can be simplified to 
\begin{align}
	\int_0^{T_{II}}d\tau \sin[2\Phi(\tau)] = 0. \label{Eq:PulseShapeII}
\end{align}

Before we proceed to find optimized pulse shapes that cancel second-order leakage errors, we want to use the above results to understand the leakage minimum of the two solid curves in Fig.~\ref{Fig:SOCFinalError}(b), which corresponds to using the square pulse in Eq.~\eqref{Eq:SquarePulse}. 
In such a case, the correction to the wave function overlap is 
\begin{align}
\xi^{(2)}(T_{II}) &= -\dfrac{\zeta^2\tildet^2}{2\hbar^2}\int_0^{T_{II}}d\tau_1 \int_0^{T_{II}}d\tau_2\cos[2\Phi(\tau_1)-2\Phi(\tau_2)]\notag\\&=-\dfrac{\zeta^2\tildet^2}{2\Omega_0^2}= -\dfrac{\zeta^2}{2}\left(\dfrac{T_{II}}{T_{II}^{(0)}}\right)^2\simeq -\dfrac{\zeta^2}{2}\times 0.94^2, \label{Eq:OverlapSquare}
\end{align}
where $T_{II}^{(0)}$ is the gate operation time if there were no ramping errors, while $T_{II} \equiv T_{II}^{(0)} - \Delta T_{II}$ is the optimized gate operation time.  
As a result, the overall leakage error is given by 
\begin{align}
	\log_{10}\eta = \log_{10}\left[2|\xi^{(2)}(T_{II})|\right]=2\log_{10}\left[{\zeta}{T_{II}}/{T_{II}^{(0)}}\right]. 
\end{align}
Therefore, the minimum for the red solid curve and the blue solid curve in Fig.~\ref{Fig:SOCFinalError}(b) should be 
at $-3.44$ and $-4.04$, respectively, which agree well with the numerical results. 

\begin{figure}[!]
\includegraphics[scale=1]{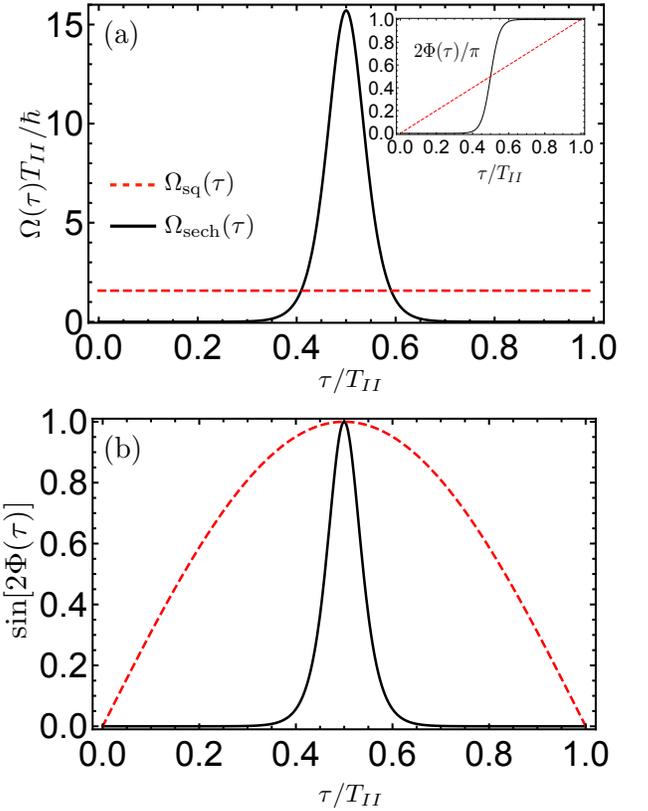}
\caption{\label{Fig:PulseShape} (a) Comparison of the square $\pi$-pulse $\Omega_\text{sq}(\tau)$ in Eq.~\eqref{Eq:SquarePulse} [dashed line] and the $\sech$-shaped $\pi$-pulse $\Omega_\text{sech}(\tau)$ in Eq.~\eqref{Eq:SechPulse} [solid line]. Here $\Omega_\text{sech}(\tau)$ has $a=20$. The inset shows a comparison of the corresponding $\Phi(\tau)$ for the two pulses. 
(b) Comparison of the integral kernel in Eq.~\eqref{Eq:PulseShapeII} for $\Omega_\text{sq}(\tau)$ and $\Omega_\text{sech}(\tau)$, respectively. }
\end{figure}

Having understood the leakage errors under a square $\pi$-pulse, we now proceed to determine which pulse shapes can satisfy the condition in Eq.~\eqref{Eq:PulseShapeII}, so that the leakage error at this order can also be minimized. 
We propose that in the $a\to \infty$ limit the following pulse shape 
\begin{align}
\Omega_\text{sech}(\tau) = \dfrac{\pi\hbar a}{4T_{II}}\sech^2\left[a\left(\dfrac{\tau}{T_{II}}-\dfrac{1}{2}\right) \right]  \label{Eq:SechPulse}
\end{align}
will be a good candidate, which is compared with the square pulse in Fig.~\ref{Fig:PulseShape}. In particular, we find that for such a pulse with $a=20$ the resulting correction to the wave function overlap is 
\begin{align}
	\xi^{(2)}(T_{II}) \simeq -\dfrac{\zeta^2}{2}\times 0.02, 
\end{align}
which is almost $45$ times smaller than that for a square pulse [cf. Eq.~\eqref{Eq:OverlapSquare}]. 
The two dashed lines in Fig.~\ref{Fig:SOCRampingError}(b) shows the overall leakage error when using this pulse with $a=20$ to flip the electron spin in stage II. We find that it indeed leads to a much reduced leakage error.

\subsection{Summary and generalization of the error-reduction scheme \label{Section:GeneralPerturbationTheory}}
In the above discussions we have developed a theory to analyze and reduce the leakage errors when transporting a single-spin qubit through a DQD with finite spin-orbit couplings. 
Although the theory was developed using the example of transporting a spin-up or spin-down state in the presence of a $\pi$-pulse, it is fairly general and can be easily extended to analyze the leakage during many other spin operations. Therefore here we would like to summarize the above formalism and extend it to incorporate a more general spin Hamiltonian. 

The Hamiltonian we consider takes the form $H_\text{eff}(\tau) = H_0(\tau) + V$, where $V = \zeta\tildet \sigma_x \otimes s_x$ is still the term that captures the spin-dependent tunnelings, and 
\begin{align}
\!\!\!\!
	H_0(\tau) =\tildet \sigma_x\otimes s_0 + \sigma_0\otimes H_s(\tau), \label{Eq:SpinHamiltonian}
\end{align}
where $H_s(\tau)$ is now a general spin Hamiltonian. 
In order for the above error correction scheme to work, we need to first identify the zeroth-order operation $U_0(\tau)$, which can perform the desired operations exactly within time $T_{II}$. For the Hamiltonian in Eq.~\eqref{Eq:SpinHamiltonian}, the spin and orbital degrees of freedom in $U_0(\tau)$ are still separable, and thus we can write 
\begin{align}
U_0(\tau) = \exp\left[-i\tildet \tau\sigma_x\otimes s_0\right]\cdot\sigma_0\otimes U_s^{(0)}(\tau), \label{Eq:Us2}
\end{align}
where $U_s^{(0)}(\tau)$ is the time-evolution operator for $H_s(\tau)$. Because $U_s^{(0)}(\tau)$ performs the desired spin rotation exactly, we must have $\bra{\chi_f} =\bra{\chi_i} U_s^{(0)}(T_{II}) $, where $\ket{\chi_{i}}$ and $\ket{\chi_f}$ are the initial and final spin states, respectively. 

We can then write the operator $V$ in the interaction picture as follows, 
\begin{align}
\hat{V}(\tau) = U_0(\tau)^\dagger V U_0(\tau) = \zeta \tildet \sigma_x \otimes \hat{V}_s(\tau),  
\end{align}
where $\hat{V}_s(\tau)$ acts only on the spin degrees of freedom. Subsequently, we can write down the $n$th order (in $\zeta$) \emph{spin} time-evolution operator in the interaction picture as follows, 
\begin{align}
\!\!\!\!
	\hat{U}_s^{(n)}(\tau) = \dfrac{1}{i^n}\dfrac{1}{n!}\int_0^\tau d\tau_1\dots \int_0^{\tau} d\tau_n \mathcal{T}\left[\hat{V}_s(\tau_1)\dots\hat{V}_s(\tau_n)\right], 
\end{align}
where $\mathcal{T}$ is the time ordering operator~\cite{bruus2004many}. As a result, the $n$th order correction to the wave function overlap is 
\begin{align}
\xi^{(n)}(T_{II})\propto \bra{\chi_f}\hat{U}_s^{(0)}(T_{II})\hat{U}_s^{(n)}(T_{II})\ket{\chi_i} = \bra{\chi_i}\hat{U}_s^{(n)}(T_{II})\ket{\chi_i}, \notag
\end{align}
where $\ket{\chi_i}$ and $\ket{\chi_f}$ are the initial and final \emph{spin} state, respectively. 
Therefore, in order to cancel out the $n$th order error, i.e., $\xi^{(n)}(T_{II})=0$, we need $\hat{U}_s^{(n)}(T_{II})$ to rotate the initial spin state $\ket{\chi_i}$ to a state that is orthogonal to it. 
This result makes it easy to read off constraints on the pulse from the form of $\hat{U}_s^{(n)}(T_{II})$: if the Bloch vector associated with the initial state $\ket{\chi_i}$ points along a direction $\hat{k}$, then we want $\hat{U}_s^{(n)}(T_{II}) \propto {\hat{p}}\cdot\bm{s}$, where $\hat{p}$ is orthogonal to $\hat{k}$, and $\bm{s} = \{s_x, s_y, s_z\}$. 
In the example we discussed in Section~\ref{Section:SpinUp}, the initial state is a spin-up state, corresponding to a Bloch vector pointing along the $\hat{z}$ direction. As a result, if the time-evolution operator has the form ${\hat{U}}_s^{(n)}(T_{II}) = a_x s_x + a_y s_y$ for any $a_x$ and $a_y$, the corresponding leakage error will vanish. 
We therefore obtain the pulse shaping constraints that the identity and $s_z$ components of ${\hat{U}}_s^{(n)}(T_{II})$ must vanish. These are exactly the implications of the two error-reduction conditions in Eq.~\eqref{Eq:PulseShapeI} and~\eqref{Eq:PulseShapeII}.

In what follows we will use this generalized formalism to analyze a few interesting scenarios, and demonstrate how it can help one derive the appropriate pulse shaping conditions for error reductions. 
For simplicity we will neglect the ramping errors for all subsequent discussions, and just focus on the errors that arise during the transport process. 

\subsection{Leakage in transporting an arbitrary spin state}

In this subsection we would like to discuss the consequences of transporting an arbitrary spin state in this spin-dependent tunneling model. 
Specifically we assume that its initial orbital wave function is an exact FD state on the left dot, while its initial spin state $\ket{\chi_{\alpha,\gamma}}$ is given generally by 
\begin{align}
	\ket{\chi_{\alpha,\gamma}} = \cos\dfrac{\alpha}{2}e^{i\gamma}\ket{\uparrow} + \sin\dfrac{\alpha}{2}\ket{\downarrow}, \label{Eq:SpinState}
\end{align}
with $\alpha \in [0, \pi]$, and $\gamma$ is some arbitrary phase of the spin state. 
In particular, $\alpha=0$ and $\pi$ correspond to the spin-up and spin-down state, respectively, while $\alpha=\pi/2$ corresponds to an eigenstate of $s_x$. 

\begin{figure}[!]
\includegraphics[scale=1]{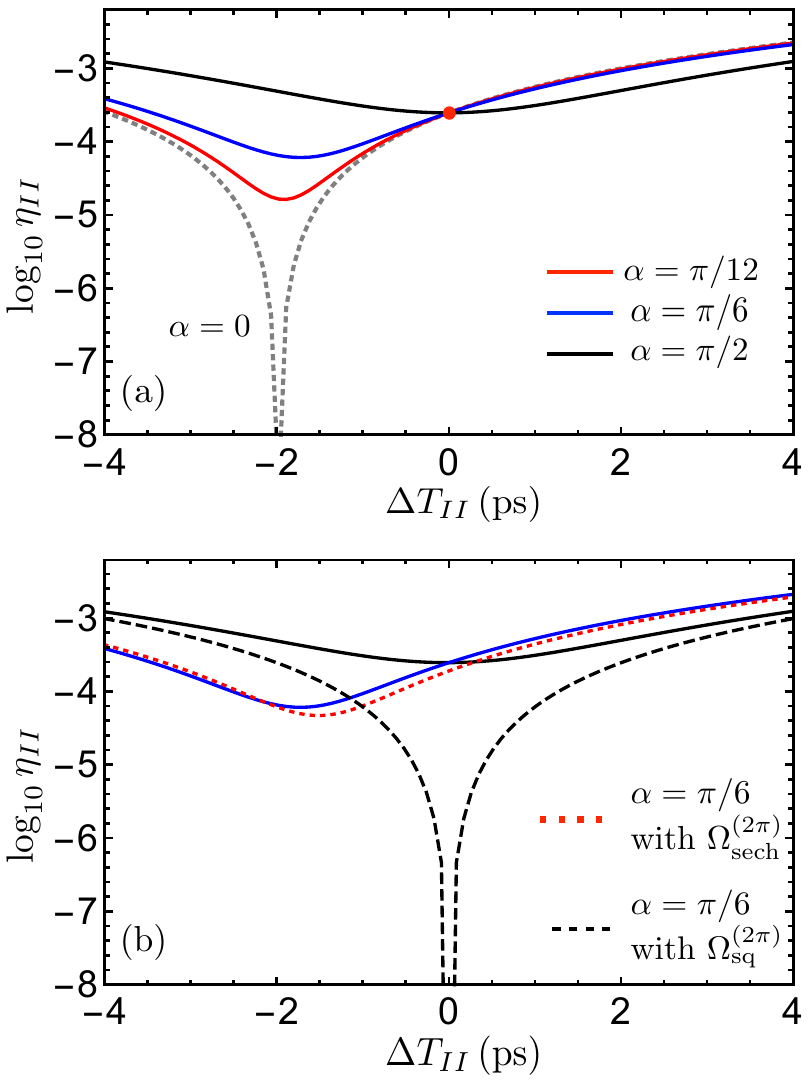}
\caption{\label{Fig:GeneralSpinError} 
(a) Error in transporting an arbitrary spin state $\ket{\chi_{\alpha}}$. This plot does not include ramping errors. The SOC parameter is set to $\zeta=0.01$, and no pulse $\Omega(t)$ is applied during the transport. 
(b) The transport error when a $2\pi$-pulse is applied to help preserve the spin state. The blue and black solid lines are the same as those in (a). 
The two pulses $\Omega_\text{sq}^{(2\pi)}$ and $\Omega_\text{sech}^{(2\pi)}$ are defined in Eq.~\eqref{Eq:SquarePulse2pi} and~\eqref{Eq:SechPulse2pi}, respectively, and the $\Omega_\text{sech}^{(2\pi)}$-pulse has a parameter of $a=20$. Also, we set the phase of all initial spin states to be zero in this plot ($\gamma=0$). 
}
\end{figure}

We find that even the simple transport of a spin state other than spin-up or spin-down is no longer exact in this model. Such an effect is illustrated in Fig.~\ref{Fig:GeneralSpinError}(a), as we now explain in details. 
First, remember from the previous discussions that transporting a spin-up state is subject to a larger tunneling rate $\tildet(1+\zeta)$ than that in a pristine system. As a result, we expect a reduced $T_{II}$ time. However, as long as we tune $\Delta T_{II}$ accordingly, such an operation is still exact, as illustrated by the gray dotted line in Fig.~\ref{Fig:GeneralSpinError}(a). 
In contrast, the transport of any other spin state will be accompanied by some finite errors. Even though one can tune $\Delta T_{II}$ to minimize such an error, the residual error is still rather large. Besides, the residual error increases with $\alpha$ monotonically for $0<\alpha<\pi/2$, and reaches its maximum at $\alpha=\pi/2$, which is marked by the red dot in Fig.~\ref{Fig:GeneralSpinError}(a). In fact we can show that the error at this red dot is first-order in $\zeta$. 
Finally, we note that in the absence of an applied pulse $\Omega(\tau)$, the leakage error will not be affected by the phase $\gamma$ of the initial spin state, because the wave function overlap here only contains expectation values of $s_0$ and $s_z$. 

From the above discussions one might worry that the potential error in transporting an arbitrary spin state is so large that it is no longer viable. Fortunately, this is not the case: such errors can be largely suppressed by applying a carefully designed $2\pi$-pulse during the transport. 
Specifically, we find that (i) the gate operation time need not be changed: one can still carry out the transport within the same $T_{II}$ time as that in a pristine system; 
(ii) in order to eliminate the first-order error in $\zeta$, the applied $2\pi$-pulse should satisfy 
\begin{align}
	\mathcal{C}_1\cos\alpha = \mathcal{S}_1\sin\alpha\sin\gamma, \label{Eq:PulseShape2piI}
\end{align}
where 
\begin{align}
	\mathcal{C}_1 = \int_0^{T_{II}} d\tau \cos[2\Phi(\tau)], \quad
	\mathcal{S}_1 = \int_0^{T_{II}} d\tau \sin[2\Phi(\tau)];  \label{Eq:C1S1}
\end{align}
(iii) in order to eliminate the second-order error in $\zeta$, the applied $2\pi$-pulse should satisfy additionally 
\begin{align}
\mathcal{C}_2 + i\mathcal{S}_2\sin\alpha\cos\gamma = 0, \label{Eq:PulseShape2piII}
\end{align}
where 
\allowdisplaybreaks[4]
\begin{align}
	\mathcal{C}_2 = \int_0^{\tau}d\tau_1 &\int_0^{\tau_1}d\tau_2 \cos[2\Phi(\tau_1)-2\Phi(\tau_2)], \notag\\
	\mathcal{S}_2 = \int_0^{\tau}d\tau_1 &\int_0^{\tau_1}d\tau_2 \sin[2\Phi(\tau_1)-2\Phi(\tau_2)]. 
\end{align}
The single equation in Eq.~\eqref{Eq:PulseShape2piII} contains two separate requirements: that $\mathcal{C}_2$ must always vanish, and that $\mathcal{S}_2 = 0$ when $\sin\alpha\cos\gamma\neq0$.
We can see from this analysis that the phase $\gamma$ of the spin state  will affect the leakage if a $2\pi$-pulse is applied to help preserve the spin state, in contrast to the case without an applied pulse. 

It is not difficult to find $2\pi$-pulses that satisfy the two conditions in Eqs.~\eqref{Eq:PulseShape2piI} and~\eqref{Eq:PulseShape2piII}. However, the set of pulses that satisfy $\mathcal{C}_1 = \mathcal{S}_1=0$ has two key advantages for spin qubit operations. First of all, the condition that $\mathcal{C}_1 = \mathcal{S}_1=0$ will immediately lead to $\mathcal{C}_2 = \mathcal{S}_2=0$. Therefore 
this set of pulses simultaneously satisfies both conditions, and thus  can eliminate errors up to at least second order in $\zeta$. More importantly, they can satisfy these two conditions regardless of the initial spin state of the spin qubit being transported. 
The fact that leakage-suppressed transport can be achieved without any knowledge of the spin state is quite surprising, and highly desirable for quantum computation and quantum information processing purposes. 

Even if we restrict our attention to pulses satisfying $\mathcal{C}_1 = \mathcal{S}_1=0$, there still remains a vast number of solutions. 
One simple and practical candidate is the following square pulse
\begin{align}
 	\Omega_\text{sq}^{(2\pi)}(\tau) = \dfrac{\pi\hbar}{T_{II}}. \label{Eq:SquarePulse2pi}
 \end{align} 
The black dashed line in Fig.~\ref{Fig:GeneralSpinError}(b) plots the error of transporting a spin state with $\alpha=\pi/6$ and $\gamma=0$ in the presence of this pulse, and indeed shows a substantial error reduction. 
In fact, the leakage error under this square pulse can be written in a closed form as follows,  
\begin{align}
\!\!\!\!
	\eta = \sin^2 \left[\pi\sqrt{1+(\zeta/2)^2}\right] 
	\left[1-\dfrac{\left(2\cos\gamma\sin\alpha+\zeta l\cos\alpha\right)^2}{4+\zeta^2}\right], \notag
\end{align}
where $l$ is the overlap between the two FD states $\psi_{L}$ and $\psi_{R}$ [cf. Eq.~\eqref{Eq:Overlap}]. The above result further shows that the leading-order error is actually proportional to $\zeta^4$, i.e., 
\begin{align}
	\eta \simeq \dfrac{\pi^2\zeta^4}{64}\left[1-\sin^2\alpha\cos^2\gamma\right]+\mathcal{O}(\zeta^5), 
\end{align}
which explains the substantial error reduction by this square $2\pi$-pulse. 
In particular, the minimum of the black dashed line in Fig.~\ref{Fig:GeneralSpinError}(b) will be at $\log_{10}\eta_{\text{min}} \simeq -8.94$. 

\begin{figure}[!]
\includegraphics[scale=1]{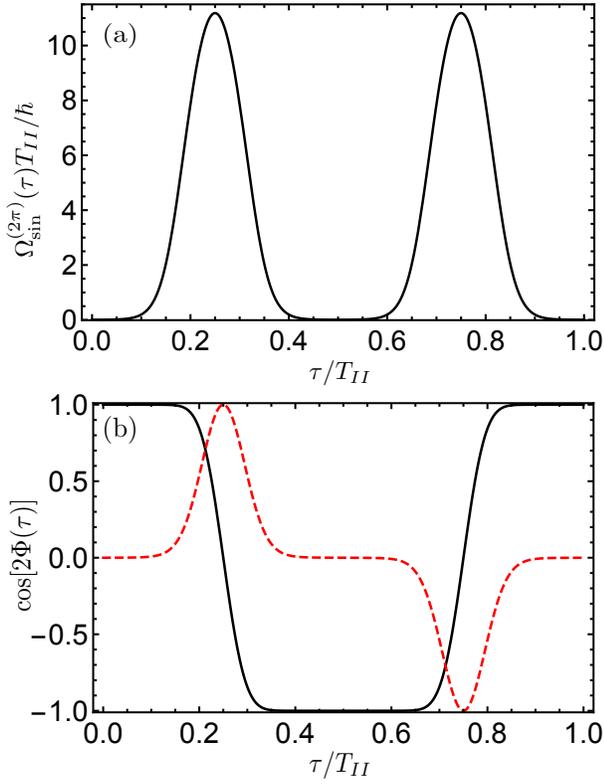}
\caption{\label{Fig:SinPulse} 
(a) Illustration of the $\Omega_\text{sin}^{(2\pi)}(\tau)$-pulse in Eq.~\eqref{Eq:SinShape}. 
(b) A plot of the corresponding $\cos[2\Phi(\tau)]$ (solid line) and $\sin[2\Phi(\tau)]$ (dashed line) for the pulse in (a). 
}
\end{figure}

We are not satisfied with just the square pulse, however, because pulses in actual experiments cannot be switched on infinitely quickly, making purely square pulses somewhat unrealistic. It is thus imperative for us to demonstrate that there exist more practical pulse shapes that satisfy the two conditions we specify in Eqs.~\eqref{Eq:PulseShape2piI} and~\eqref{Eq:PulseShape2piII}. 
Fortunately such pulses do exist. As an example of more practical pulses that still satisfy $\mathcal{C}_1 = \mathcal{S}_1=0$ we design the following pulse, 
\begin{align}
\!\!\!\!
	\Omega_\text{sin}^{(2\pi)}(\tau) = \dfrac{\hbar\mathcal{A}}{T_{II}}\dfrac{\sin^2\left[\pi(4z-1)/2\right] + 2\sin^2\left[\pi(4z-1)\right]}{\sqrt{e^{16\sin^2[\pi(4z-1)/2]} - 4\sin^2[\pi(4z-1)] - 1}}, \label{Eq:SinShape}
\end{align}
where $z = \tau/T_{II}\in (0,1)$, and $\mathcal{A} \simeq 16\pi^2/10.59$. Such a pulse is shown in Fig.~\ref{Fig:SinPulse}. 
Numerical simulations show that this pulse can preserve the initial spin state very well. In fact, the leakage error under this pulse nearly coincides with the result for the square pulse in Eq.~\eqref{Eq:SquarePulse2pi}, and thus is not separately shown in Fig.~\ref{Fig:GeneralSpinError}(b). However, we do find that the leakage minimum for this pulse is even lower, at $\log_{10}\eta_{\text{min}} \simeq -9.40$. 

Before we end this subsection, we want to make some additional comments on the two conditions~\eqref{Eq:PulseShape2piI} and~\eqref{Eq:PulseShape2piII}. 
First, we emphasize that they are in fact the most general error reduction conditions one can write down in the zero spin detuning case $(\delta=0)$, and therefore they constitute one of the main results of this paper. 
In particular, the two conditions we derived in Eq.~\eqref{Eq:PulseShapeI} and~\eqref{Eq:PulseShapeII} correspond to the special limits of $\mathcal{C}_1 = 0$ and $\mathcal{C}_2 = 0$, respectively. 

\begin{figure}[!]
\includegraphics[scale=1]{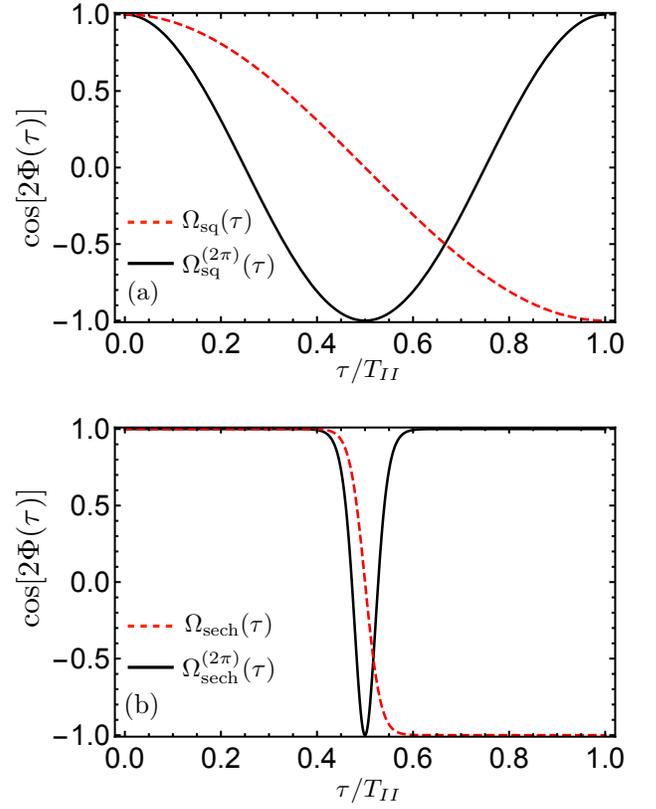}
\caption{\label{Fig:PulseShape2Pi} 
Comparison between $\pi$-pulses and $2\pi$-pulses. The solid lines correspond to $2\pi$-pulses, while the dashed lines correspond to their $\pi$-pulse counterparts. 
(a) and (b) correspond to the square pulse and $\sech$-shaped pulse, respectively. 
}
\end{figure}

In addition, one should not be deceived by the innocuous form of the two conditions in Eqs.~\eqref{Eq:PulseShape2piI} and~\eqref{Eq:PulseShape2piII}: although not obvious, these two conditions crucially depend on the area of the pulse. 
For example, when $\gamma=0$ the condition~\eqref{Eq:PulseShape2piI} reduces to $\mathcal{C}_1 = 0$, which looks almost identical to Eq.~\eqref{Eq:PulseShapeI}. 
However, this is where the similarity ends. Unlike $\pi$-pulses, symmetric $2\pi$-pulses will no longer satisfy $\mathcal{C}_1 = 0$ by default, and thus in general cannot eliminate first-order errors in transporting a spin state. Moreover, unlike its $\pi$-pulse counterpart, the following $\sech$-shaped $2\pi$-pulse,
\begin{align}
\Omega_\text{sech}^{(2\pi)}(\tau) = \dfrac{\pi\hbar a}{2T_{II}}\sech^2\left[a\left(\dfrac{\tau}{T_{II}}-\dfrac{1}{2}\right) \right],  \label{Eq:SechPulse2pi}
\end{align}
is also no longer a good candidate to preserve the spin here: the effect of this pulse is shown by the red dotted line in Fig.~\ref{Fig:GeneralSpinError}(b), and we can see that it is only barely better than not applying a pulse at all [cf. the blue line therein]. 
All these differences are attributed to the fact that $2\pi$-pulses have twice the area of a $\pi$-pulse. Figure~\ref{Fig:PulseShape2Pi} provides an intuitive understanding of why the square pulse in Eq.~\eqref{Eq:SquarePulse2pi} satisfies the condition $\mathcal{C}_1 = 0$ while the $\sech$-shaped pulse in Eq.~\eqref{Eq:SechPulse2pi} does not. 
The influence of different pulse shapes on spin shuttling is thus quite subtle and nonintuitive.

\subsection{Leakage of an arbitrary rotation by $s_x$}

\begin{figure}[!]
\includegraphics[scale=1]{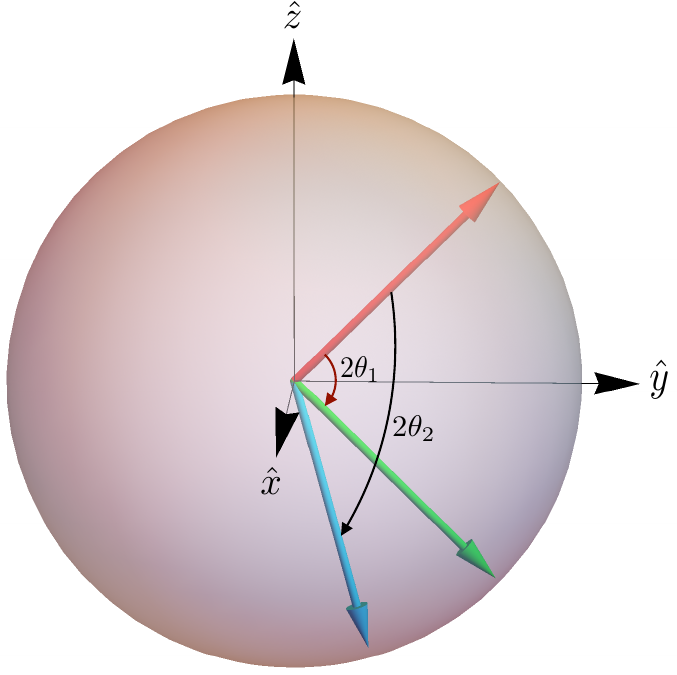}
\caption{\label{Fig:Sphere} 
Illustration of two different $s_x$ rotations. The initial spin state is $\alpha = \pi/4$ and $\gamma=\pi/2$, which is marked by the red arrow. The angle of the two rotations are $2\theta_1 = \pi/2$ and $2\theta_2 = 2\pi/3$, respectively. 
}
\end{figure}

Having discussed the transport of an arbitrary initial spin state, one may wonder what the consequences are of flipping an arbitrary spin state by a $\pi$-pulse during the transport. More generally, one can even consider the consequences of performing a general $s_x$-rotation on an arbitrary spin state during the transport. 
In this section we would like to address these questions, which will also provide a good illustration of how to make use of the general pulse shaping conditions in Eqs.~\eqref{Eq:PulseShape2piI} and~\eqref{Eq:PulseShape2piII}.

The question we want to address is illustrated in Fig.~\ref{Fig:Sphere}. Specifically, we study the leakage when an arbitrary spin state (red arrow) undergoes a general $s_x$-rotation during the transport. The angle of the spin rotation will be $2\theta$, with $\theta$ being the area of the pulse that performs the $s_x$ rotations. 
To simplify our discussions we will further assume that $\gamma\equiv\pi/2$, so that the spin state always lies in the $y$-$z$ plane. 
Thus, the first-order error reduction condition in Eq.~\eqref{Eq:PulseShape2piI} simplifies to 
\begin{align}
  \mathcal{C}_1 \cos\alpha = \mathcal{S}_1\sin\alpha. \label{Eq:PulseShapeTheta-I}
\end{align}
One then needs to determine what pulse shapes can satisfy this condition and thereby eliminate first-order leakage errors. 
Note that for this problem it is generally no longer possible to find pulse shapes that suit all initial spin states and/or all different $s_x$ rotations; instead we have to tailor the pulse shapes for each individual case. The reason is that in many cases the $s_x$ rotation performed simply precludes the possibility that $\mathcal{C}_1 = \mathcal{S}_1 = 0$. For example, for an $s_x$ rotation by a pulse with area $\pi/3$, the integrand of $\mathcal{S}_1$ is always positive, and thus no pulse shapes can possibly satisfy $\mathcal{S}_1 = 0$. 

In the following we present two examples to illustrate how the condition in Eq.~\eqref{Eq:PulseShapeTheta-I} can help one design appropriate pulse shapes for efficient error corrections.

As the first example, we would like to show the effects of square and $\sech$-shaped pulses, defined as
\begin{align}
 \Omega^{(\theta)}_\text{sq} = \dfrac{\theta\hbar}{T_{II}}, \quad 
 \Omega_\text{sech}^{(\theta)}(\tau) = \dfrac{\theta\hbar a}{2T_{II}}\sech^2\left[a\left(\dfrac{\tau}{T_{II}}-\dfrac{1}{2}\right) \right]. \label{Eq:PulseTheta}
\end{align}
They both cover an area of $\theta$ for $\tau\in(0,T_{II})$, and thus can perform an $s_x$-rotation with an angle of $2\theta$. 
However, these two pulses are limited in the sense that the corresponding  ratio $\mathcal{C}_1/\mathcal{S}_1$ is fixed by their area, i.e., $\mathcal{C}_1/\mathcal{S}_1 \equiv \cot\theta$. As a result, they can only satisfy the constraint in Eq.~\eqref{Eq:PulseShapeTheta-I} for $\alpha = \pi/2-\theta$. Therefore, they will only be useful for the specific operation that the initial spin state $\ket{\chi_{\alpha,\pi/2}}$ is rotated about the $x$-axis by an angle of $\pi-2\alpha$. Such a rotation for the initial state of $\alpha=\pi/4$ is illustrated by the rotation marked by $2\theta_1$ in Fig.~\ref{Fig:Sphere}. 

For more general $s_x$ rotations one has to find in each case the pulses for which the ratio $\mathcal{C}_1/\mathcal{S}_1$ can be adjusted independently of their area. For instance, let us consider 
\begin{align}
	\Omega_\text{cos}(\tau) = \dfrac{\pi \hbar a\theta}{2T_{II}}\left(\dfrac{\tau}{T_{II}}\right)^{a-1} \cos\left[\dfrac{\pi}{2}\left(\dfrac{\tau}{T_{II}}\right)^a\right], \; (a>1) \label{Eq:PulseShapeCosine}
\end{align}
which covers a fixed area $\theta$ regardless of $a$. Therefore, we can use $a$ to adjust the ratio $\mathcal{C}_1/\mathcal{S}_1$, and adapt the pulse shape  for a given initial spin state $\alpha$. As an example, if we want to rotate the same initial spin state $\alpha = \pi/4$ in Fig.~\ref{Fig:Sphere} by a different angle $2\theta_2 =2\pi/3$, we can use this pulse with $a\approx 2.379$, so that at least the first-order leakage error can be eliminated.

\subsection{Generalization to arbitrary spin rotations}
Having discussed the transport and $s_x$-rotation of an arbitrary spin state, we would like to finally generalize to also include arbitrary spin rotations. To achieve this we need to drop the assumption that the driving field $\Omega(t)$ is always resonant with the Zeeman splitting. Therefore, the Hamiltonian we consider in this subsection is given by $H_\text{eff}(\tau) = H_0(\tau) + V$, where $V = \zeta\tildet \sigma_x \otimes s_x$ and 
\begin{align}
\!\!\!\!
	H_0(\tau) =\tildet \sigma_x\otimes s_0 + \sigma_0\otimes H_s(\tau),\, H_s(\tau) = \Omega(\tau) s_z +\dfrac{\delta}{2}s_x, \label{Eq:SpinHamiltonian-2}
\end{align}
where $\delta$ is the spin detuning. Note that for convenience the above Hamiltonian has been written in a different spin basis than that in Eq.~\eqref{Eq:Hamiltonian-1}, and that they are related by a Hadamard transformation,  
\begin{align}
 \mathcal{U} = \dfrac{1}{\sqrt{2}}\TwoDMatrix{1}{1}{1}{-1}. 
\end{align}

To solve this time-dependent problem, we first note that the two terms in $H_0(\tau)$ still commute with each other. As a result, the time-evolution operator for $H_0(\tau)$ are still given generally by Eq.~\eqref{Eq:Us2}.  
However, its spin part $U_s^{(0)}(\tau)$ no longer assumes the simple form in Eq.~\eqref{Eq:Us1}, and thus needs to be calculated separately for each different spin operations. 
Nonetheless, some general properties of $U_s^{(0)}(\tau)$ can still be drawn. We start by noting that because $U_s^{(0)}(\tau)$ is unitary, we can parametrize it generally as follows~\cite{Barnes2012:Analytical,Barnes2015:SpinCubit}
\begin{align}
U_s^{(0)}(\tau) = \TwoDMatrix{u_{11}}{-u_{21}^{\ast}}{u_{21}}{u_{11}^{\ast}}, 
\end{align}
where both $u_{11}$ and $u_{21}$ are ordinary complex numbers. If we introduce a pair of new variables 
\begin{align}
	D_{\pm} = \dfrac{1}{\sqrt{2}} e^{\pm i\delta \tau/2} (u_{11}\pm u_{21}), 
\end{align}
the Schroedinger equation for $u_{11}$ and $u_{21}$ can be written compactly as follows, 
\begin{align}
	\dot{D}_{\pm} = -i\Omega(\tau) e^{\pm i\delta \tau} D_{\mp}. \label{Eq:SchroedingerEquation}
\end{align}
We can then write down the spin-orbit coupling term $V$ in the interaction picture as follows, 
\begin{align}
\hat{V}_s(\tau) = \TwoDMatrix{|D_{+}|^2-|D_{-}|^2}{2D_{+}^\ast D_{-}^{\ast}}{2D_{+} D_{-}}{|D_{-}|^2-|D_{+}|^2}. 
\end{align}
Thus, according to the general formalism in Section~\ref{Section:GeneralPerturbationTheory}, the condition to eliminate the first-order error is given by
\begin{align}
	\bra{\tilde{\chi}_i} \left[\int_0^{T_{II}}d\tau_1 \hat{V}_s(\tau_1)\right] \ket{\tilde{\chi}_i} = 0, \label{Eq:63}
\end{align}
while the condition to eliminate the second-order error is
\begin{align}
\bra{\tilde{\chi}_i} \left[\int_0^{\tau}d\tau_1\int_0^{\tau_1}d\tau_2 \hat{V}_s(\tau_1)\hat{V}_s(\tau_2)\right] \ket{\tilde{\chi}_i} = 0. \label{Eq:64}
\end{align}

The condition in Eq.~\eqref{Eq:63} can be written more explicitly. There is one caveat, however, as we have transformed to a different spin basis. As a result, the initial spin state here $\ket{\tilde{\chi}_i}$ is related to the more familiar one in Eq.~\eqref{Eq:SpinState} by 
\begin{align}
	\ket{\tilde{\chi}_i} = \mathcal{U}\ket{\chi_{\alpha,\gamma}}. 
\end{align}
After some simplifications, the first-order error-reduction condition in Eq.~\eqref{Eq:63} reads, 
\begin{align}
	\mathcal{A}\sin\alpha\cos\gamma + 2\text{Re}[\mathcal{B}]\cos\alpha + 2\text{Im}[\mathcal{B}]\sin\alpha\sin\gamma = 0, 
\end{align}
with 
\begin{align}
	\mathcal{A} = \int_0^{T_{II}} d\tau_1 \left[|D_{+}|^2-|D_{-}|^2\right],\; 
	\mathcal{B} = \int_0^{T_{II}} d\tau_1 D_{+}D_{-}. 
\end{align}
The further simplification of Eq.~\eqref{Eq:64} is less useful, and we will not write it down here. 

In general, the two conditions~\eqref{Eq:63} and~\eqref{Eq:64} impose constraints on the functional form of $D_+$ and $D_-$, which in turn place requirements on the pulse shape $\Omega(\tau)$. 
However, the general solution of this problem is beyond the scope of this work, and we will leave it for the future. 

\section{Leakage in spin qubits based on a single silicon quantum dot \label{Section:Silicon}}

Having discussed leakage errors due to finite spin-orbit couplings, we now turn to the last source of leakage errors: the presence of multiple valley states, (e.g., in Si or Ge conduction band). 
In semiconductors it is common that there exist more than one minimum in the conduction band, and that they are degenerate with each other. These degenerate conduction band minima are called the ``valleys'' of the band structure, and they often give rise to interesting electronic properties~\cite{LiXiao2013:MoS2,LiXiao2014:DomainWall,LiXiao2015:SmB6,LiXiao2016:SnTe}. %

Silicon, one of the most-studied spin quantum computing platforms, is well known for its multivalley structure in the conduction band. 
In particular, there are six valleys in its bulk conduction band. Upon confinement along the growth direction $\hat{z}$, the six valleys split into a low-energy twofold-degenerate subspace, and a high-energy fourfold-degenerate one in the two-dimensional quantum dot systems of interest in the present work. 
The energy separation between these two valley subspaces is of the order of a few meV~\cite{DasSarma1979:stress}, and thus the leakage into the high-energy subspace is highly suppressed, and will not likely pose a problem for spin-qubit operations within the low-energy subspace. 
However, the question still remains of whether the presence of a second valley state in the low-energy subspace will complicate the operation of a single-spin qubit.

In this section we will analyze this problem in some detail. 
However, our discussions will be limited to the leakage in operating single-spin qubits in a \emph{single} silicon quantum dot, for several reasons. 
First, the leakage due to multiple valley states already fully manifests itself in a single silicon quantum dot. Therefore, it seems appropriate to address the single-dot case first.  
Second, the single-dot double-valley structure in silicon makes a nice comparison with the double-dot single-valley structure in GaAs, which is interesting in its own right. 
Third, in certain limits the generalization to double quantum dots can be straightforward, especially because we only consider one electron problems~\footnote{The situation can become more involved for two-electron problems due to electron-electron interactions.}. 
For example, if we assume that the two quantum dots have identical valley-orbit couplings (defined below), the electron will only tunnel between like valley eigenstates~\cite{Culcer2010:SiliconQubit}. 
In this case many results in the double quantum dot in Si case will be qualitatively the same as those we found for GaAs. 
We emphasize that now we are considering two valleys and one quantum dot, whereas before we considered two quantum dots and one valley.  Later, we will compare and contrast these situations. 

The main results of this section can be summarized as follows. 
(i) a rough interface potential will couple the twofold degenerate valley states at $\bm{k}_{\pm} = \pm k_0\hat{z}$ (this mechanism is often referred to as the ``valley-orbit coupling''), leading to a valley splitting $\Delta$ between them; 
(ii) in the limit of $\Delta\gg k_BT$, a single-spin qubit can be initialized with a high fidelity within the two spin states in the lower valley eigenstate; 
(iii) in a pristine system, the two valley eigenstates are orthogonal to each other, and there will be no leakage to the other valley eigenstate, regardless of their energy separations; 
(iv) external perturbations like spin-orbit couplings will indirectly couple the two valley eigenstates, leading to finite leakage errors near certain ``leakage hot spots''. However, such leakage errors can be avoided if the gate operations are performed away from such leakage hot spots. 
We now provide below the theoretical details leading to these conclusions.

\subsection{Valley eigenstates in a single silicon quantum dot}
We consider a single silicon quantum dot located at $\bm{R} = (X_0, 0, 0)$, described by the following Hamiltonian 
\begin{align}
	H = T_0 + V(\br) + H_v, \label{Eq:SiliconHamiltonian}
\end{align}
where $T_0$ and $V(\br)$ are the kinetic energy and the confinement potential respectively, given by 
\begin{align}
\!\!\!\!
	T_0 = \dfrac{(\bm{p}+e\bm{A})^2}{2m_t}, V(\br) = \dfrac{\hbar\omega_0}{2a^2}\left[(x-X_0)^2+y^2\right] + \dfrac{\hbar\omega_z}{2b^2}z^2, 
\end{align}
where $m_t=0.191m_0$ is the transverse effective mass for Silicon ($m_0$ is the bare electron mass). Moreover, $a=\sqrt{\hbar/m_t\omega_0}$ is the in-plane FD radius, and $b$ is the growth-direction confinement length. The term $H_v$ in the Hamiltonian is the valley-orbit coupling term which we will discuss below~\footnote{Note that we consider only the ground two valley states since the upper valleys are $\sim\SI{10}{meV}$ above the ground state, and can be safely ignored for our considerations.}. 

In the absence of $H_v$, the two valley states in this low-energy branch remain degenerate, and their wave functions can be written generally as
\begin{align}
	\psi_{\lambda}(\br) = \varphi(\br) e^{i\bk_{\lambda}\cdot\br}u_{\lambda}(\br), \label{Eq:BareValleyStates}
\end{align}
where $\lambda = \pm$ labels the two valleys located at $\bk_{\pm} = \pm k_0\hat{z}$, with $k_0 = 0.85(2\pi/\asi)$, and the lattice constant of silicon is $\asi = \SI{5.43}{\AA}$. 
The envelope function is 
\begin{align}
	\varphi(\br) = \dfrac{1}{\pi^{3/4}\sqrt{a^2 b}} e^{-[(x-X_0)^2+y^2]/2a^2}e^{-z^2/2b^2}, 
\end{align}
while the lattice-periodic part is $u_{\lambda}(\br) = \sum_{\bK} c_{\lambda,\bK} e^{i\bK\cdot\br}$, with $\bK$ the reciprocal lattice vectors.

The overlap between the two valley states $\ket{\psi_{\pm}}$ is nonzero, but is suppressed by an exponential factor $e^{-{b^2} Q_z^2/4}$, with $Q_z =({2\pi \delta n_z}/{\asi})-2k_0$, and $\delta n_z$ is an integer. 
With a typical confinement length $b\simeq \SI{3}{nm}$~\cite{Culcer2010:Cubit}, we find that $b^2Q_z^2/4 \simeq (29.5)^2 = 870.25$ for $\delta n_z=0$, and that $b^2Q_z^2/4 \simeq (12.2)^2 = 148.84$ for $\delta n_z=1$. However, for a smaller $b$, this overlap can be larger. 
Note that the length parameter $b$ is nonuniversal and depends on the theoretical details of the confinement potential in the $\hat{z}$-direction~\cite{DasSarma1979:stress,Stern1984electron}, except that typically $b\ll X_0$.

\subsubsection{Valley-orbit coupling and valley eigenstates}
The degeneracy of the valley states will be lifted if the interface potential is sharp on the atomic scale, leading to a finite valley splitting $|\Delta|$, which can be as large as $\SI{0.8}{meV}$, but is typically $\sim \SI{0.1}{meV}$~\cite{Boykin2004:ValleySplitting,Li2010:Qubits,Culcer2009:SpinQubit,Saraiva2009:SpinQubit,Culcer2010:SiliconQubit,Friesen2010:valley-orbit,Culcer2010:Cubit,Saraiva2011:SpinQubit,Yang2013:SiliconCubit}. Moreover, both the magnitude and phase of $\Delta$ can be controlled by an electric field in the experiment. 
In this work we adopt a phenomenological model~\cite{Culcer2010:SiliconQubit} for the valley-orbit coupling term $H_v$ in the Hamiltonian~\eqref{Eq:SiliconHamiltonian}, with 
\begin{align}
	\bra{\tilde{\psi}_{+}}H_v\ket{\tilde{\psi}_{-}} =|\Delta|e^{-i\varphi}. \label{Eq:ValleyOrbit}
\end{align}
Note that similar to the case of double quantum dots, we have introduced a pair of orthogonalized valley states $\ket{\tilde{\psi}_{\pm}}$, which are related to the bare valley states in Eq.~\eqref{Eq:BareValleyStates} as follows, 
\begin{align}
	\ket{\tilde{\psi}_{+}} = \dfrac{\ket{\psi_{+}}-g\ket{\psi_{-}}}{\sqrt{1-2gl+g^2}}, \quad 
	\ket{\tilde{\psi}_{-}} = \dfrac{\ket{\psi_{-}}-g\ket{\psi_{+}}}{\sqrt{1-2gl+g^2}}, 
\end{align}
where $l = \neiji{\psi_{+}}{\psi_{-}}$ is the overlap between the bare valley states, and $g = l/(1+\sqrt{1-l^2})$. 
We can then diagonalize $H_v$ in this basis, and obtain the \emph{valley eigenstates} $(\tilde{\psi}_{+}\pm e^{i\varphi} \tilde{\psi}_{-})/\sqrt{2}$. These two valley eigenstates are orthogonal, and are separated by an energy $\EVS=2|\Delta|$, which is a parameter of our theory.

\subsection{Leakage in a single-spin qubit in silicon}
We would like to examine the role of the valley degree of freedom in the operation of a single-spin qubit in silicon; in particular, whether the two closely spaced valley eigenstates will give rise to serious experimental challenges in spin transport. 
One important issue is the proper initialization of the single-spin qubit, which depends crucially on the size of the valley splitting $2|\Delta|$: only in the limit of $2|\Delta|\gg k_BT$ can one initialize the electron into a definite orbital state with a high fidelity~\cite{Culcer2009:SpinQubit,Culcer2010:Cubit,Kawakami2014}. 
Given that the experimental value of Si valley splitting is $\SI{0.1}{meV}$ or larger, and the experiments are typically carried out at less than $\SI{100}{mK}$, the low temperature condition applies, and we consider only the $T=0$ situation.

Our main interest in this work, however, is in whether the nearby valley state will lead to additional leakage channels in the spin qubit operations. In a pristine system the answer is obviously no, because as we showed above, the two valley eigenstates are orthogonal to each other. As a result, single spin rotations in a given valley eigenstate can be performed with no leakage to the other valley, as long as there is a finite splitting between the two valleys. 
We will not consider the case where the two valleys are degenerate, because in such a case we cannot define a single-spin qubit to begin with, and Si is then (for zero valley splitting) no longer a good candidate for spin quantum computing purposes.

However, if some external perturbations couple the two valley eigenstates, the leakage to the other valley state during the spin rotations cannot be neglected. Here we use spin-orbit coupling as a possible mechanism for the intervalley coupling and present quantitative estimates of leakage in such a situation. 
Obviously, the combined presence of both valley-orbit and spin-orbit couplings will lead to some leakage errors associated with the second valley state, since the pure orthogonality of the pristine valley states no longer applies. 

\subsubsection{Intervalley coupling induced by spin-orbit couplings}
\begin{figure}[!]
\includegraphics[scale=1]{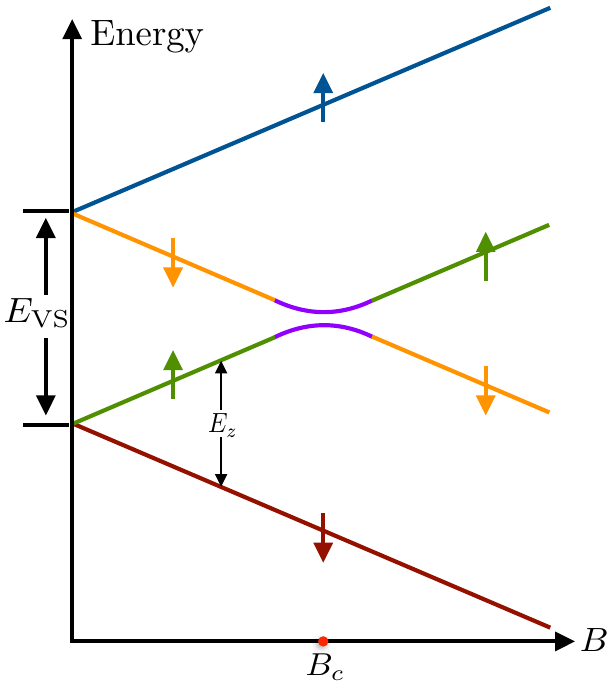}
\caption{\label{Fig:EnergyDiagram} Illustration of the valley and Zeeman splitting in a single-spin qubit in silicon. The red dot labeled by $B_c$ marks the ``hot-spot'' where the Zeeman splitting equals the valley splitting. }
\end{figure}

Recent experiments~\cite{Yang2013:SiliconCubit} have shown that an effective coupling between the two valley eigenstates $\tilde{\psi}_{\pm}$ can be induced by spin-orbit couplings at the interface of silicon quantum dots. 
These experimental results were empirically explained by an effective model to demonstrate that the spin relaxation rate is dramatically enhanced near a ``hot-spot'', defined as the resonant point where the valley splitting coincides with the Zeeman splitting, as shown in Fig.~\ref{Fig:EnergyDiagram}. 
Here we develop a theory based on this empirical model to estimate the leakage during the gate operations in the presence of valley-orbit and  spin-orbit couplings. We will show that the leakage will also become pronounced near the hot-spot, and that one needs to avoid such hot-spots in order to minimize the leakage during Si (and other multi-valley semiconductor) gate operations. 

We will start by reviewing the model constructed in Ref.~\onlinecite{Yang2013:SiliconCubit}. 
The spin-orbit Hamiltonian is still given by the general form in Eq.~\eqref{Eq:Spin-OrbitHamiltonian}. 
In the experiment the confinement of the quantum dot is along the $(001)$ (i.e., $\hat{z}$) direction, while the magnetic field is applied along the $(110)$ direction. The corresponding vector potential is $\bm{A} = \frac{B}{2\sqrt{2}}(z, -z, -x + y)$. 
As a direct consequence, the wave function of the quantum dot states will not be modified by the applied in-plane magnetic field. Furthermore, the definition of spin-up and spin-down states is changed from the eigenstates of $\sigma_z$ to the eigenstates of the following operator, 
\begin{align}
	\sigma_{+45} \equiv -(\sigma_x + \sigma_y), \quad 
	\sigma_{+45}\ket{\uparrow, \downarrow} = \pm \ket{\uparrow, \downarrow}. 
\end{align}

Because the matrix element of the momentum operator can be directly mapped onto dipole moments in this model~\cite{Yang2013:SiliconCubit}, 
\begin{align}
	\dfrac{i\hbar}{m_t} \bra{\tilde{\psi}_{-}} P_x \ket{\tilde{\psi}_{+}} = \EVS \bra{\tilde{\psi}_{-}} \hat{x} \ket{\tilde{\psi}_{+}}, 
\end{align}
it is then convenient to express the spin-orbit Hamiltonian in Eq.~\eqref{Eq:Spin-OrbitHamiltonian} in terms of the dipole matrix elements between the two valley eigenstates, 
\begin{align}
\bra{\tilde{\psi}_{+}}x\ket{\tilde{\psi}_{-}} = \bra{\tilde{\psi}_{+}}y\ket{\tilde{\psi}_{-}} = r, 
\end{align}
which is treated as a phenomenological parameter of the model. 
In particular, the explicit form of this matrix element is given by~\cite{Yang2013:SiliconCubit} 
\begin{align}
\bra{\tilde{\psi}_{-,\uparrow}^{(0)}}H_\text{SO}\ket{\tilde{\psi}_{+,\downarrow}^{(0)}} = \sqrt{2}r\dfrac{m_t\EVS}{\hbar}(\beta_D-\alpha_R),  
\end{align}
where the superscript $(0)$ indicates unperturbed states. 
The parameters extracted from the experiment~\cite{Yang2013:SiliconCubit} are $\beta_{D}-\alpha_{R} \simeq \SI{15}{m/s}$, $\EVS \simeq \SI{0.75}{meV}$, and $r\simeq \SI{2}{nm}$, so that the magnitude of this intervalley coupling is about $\SI{0.05}{\micro eV}$. Other matrix elements of the spin-orbit coupling Hamiltonian are in general much smaller. For example, intervalley coupling between the same spin states is negligible, as 
\begin{align}
	\bra{\tilde{\psi}_{-, \downarrow}^{(0)}} H_{\text{SO}} \ket{\tilde{\psi}_{+,\downarrow}^{(0)}} \propto 
	\bra{\tilde{\psi}_{-}} \hat{x} \ket{\tilde{\psi}_{+}} - \bra{\tilde{\psi}_{-}} \hat{y} \ket{\tilde{\psi}_{+}} \simeq 0. 
\end{align}
Moreover, the dipole moment within the same valley is likely small, and thus we will neglect the effect of spin-orbit coupling between states in the same valley. 

The analysis above indicates that near the hot-spot $B_c$, the two states $\tilde{\psi}_{-,\uparrow}$ and $\tilde{\psi}_{+,\downarrow}$ will be renormalized as follows~\cite{Yang2013:SiliconCubit}, 
\allowdisplaybreaks[4]
\begin{align}
\tilde{\psi}_{-,\uparrow}^{(1)} &= \sqrt{\dfrac{1-\xi}{2}}\tilde{\psi}_{-,\uparrow}^{(0)}-\sqrt{\dfrac{1+\xi}{2}}\tilde{\psi}_{+,\downarrow}^{(0)},\notag\\
\tilde{\psi}_{+,\downarrow}^{(1)} &= \sqrt{\dfrac{1+\xi}{2}}\tilde{\psi}_{-,\uparrow}^{(0)}+\sqrt{\dfrac{1-\xi}{2}}\tilde{\psi}_{+,\downarrow}^{(0)}, \label{Eq:RenormalizedStates}
\end{align}
where the superscript ``$(1)$'' and ``$(0)$'' indicate renormalized and unperturbed states, respectively. In addition, 
\begin{align}
	\xi = -E_d/(E_d^2 + \Delta_a^2)^{1/2} \in (-1,1)
\end{align}
characterizes the mixing between these two states, and $E_d = \EVS-E_Z$ is the detuning from the hot-spot $B_c$. Finally, the energy splitting between $\tilde{\psi}_{-,\uparrow}^{(1)}$ and $\tilde{\psi}_{+,\downarrow}^{(1)}$ at $B_c$ is 
\begin{align}
	\Delta_a \equiv 2|\bra{\psi_{L},\uparrow} H_{\text{SO}} \ket{\psi_{H}, \downarrow}|. 
\end{align}

\subsubsection{Leakage of spin rotations near $B_c$}
We now analyze the leakage of spin rotations near the hot-spot $B_c$, and estimate how far away from $B_c$ the operation must be carried out in order for leakage errors to be negligible. 
We assume that the valley splitting is large enough that the spin qubit can be initialized with high fidelity in the two spin states in the lower valley eigenstate $\tilde{\psi}_{-}$. 
The dynamics of the spin qubit is governed by a time-dependent driving Hamiltonian $H(t)$, which can be written in the basis of $\{\tilde{\psi}_{-,\downarrow}, \tilde{\psi}_{-,\uparrow}^{(1)}, \tilde{\psi}_{+,\downarrow}^{(1)},\tilde{\psi}_{+,\uparrow}\}$ as follows, 
\begin{align}
	H(t) = 
	\begin{pmatrix}
	 -\dfrac{E_s}{2} & \Omega_{-}(t) e^{+i\omega t} & \Omega_{+}(t) e^{+i\omega t} & 0 \\
	 \Omega_{-}(t) e^{-i\omega t} &-\dfrac{\sqrt{E_d^2+\Delta_a^2}}{2}& 0 &-\Omega_{+}(t)e^{+i\omega t} \\
	 \Omega_{+}(t) e^{-i\omega t} & 0 & \dfrac{\sqrt{E_d^2+\Delta_a^2}}{2} & \Omega_{-}(t) e^{+i\omega t} \\
	 0 & -\Omega_{+}(t) e^{-i\omega t} & \Omega_{-}(t) e^{-i\omega t} & \dfrac{E_s}{2}
	\end{pmatrix}, \label{Eq:SiliconH}
\end{align}
where we have defined 
\begin{align}
	\Omega_{\pm}(t) = \sqrt{\dfrac{1\pm\xi}{2}}\Omega(t), \quad E_s \equiv \EVS+E_Z, 
\end{align}
while $E_d$ and $\Delta_a$ are defined under Eq.~\eqref{Eq:RenormalizedStates}. 
We can again introduce a unitary transformation to remove the oscillatory time dependence in Eq.~\eqref{Eq:SiliconH}, and obtain 
\begin{align}
	H_\text{eff}(t) = 
	\begin{pmatrix}
	 \omega-\dfrac{E_s}{2} & \Omega_{-}(t)  & \Omega_{+}(t)  & 0 \\
	 \Omega_{-}(t)  & -\dfrac{\sqrt{E_d^2+\Delta_a^2}}{2} & 0 & -\Omega_{+}(t)  \\
	 \Omega_{+}(t)  & 0 & \dfrac{\sqrt{E_d^2+\Delta_a^2}}{2} & \Omega_{-}(t)  \\
	 0 & -\Omega_{+}(t)  & \Omega_{-}(t)  & \dfrac{E_s}{2}-\omega
	\end{pmatrix}. \label{Eq:SiliconEffH}
\end{align}
The pulse $\Omega(t)$ is characterized by two independent parameters, i.e., its duration and the detuning of the driving $\delta$, defined as the energy difference between the driving frequency and the energy separation of the targeted transitions. 

As an example, we study the leakage during a spin rotation under a $\pi$-pulse, which can flip between spin-down and spin-up states with a $100\%$ probability in the ``fast limit'', defined as $T_0\ll (\pi\hbar)/2\delta$. 
To begin with, we initialize the electron in the lower valley eigenstate with spin-down. A $\pi$-pulse is then applied to flip its spin. Because of the couplings between the two valley subspaces, it is expected that there is a finite probability for the electron to end up in the other valley eigenstate. 
We can characterize the leakage error of such an operation as, 
\begin{align}
	\eta = 1-|\bra{\tilde{\psi}_{-,\downarrow}}U(T_0)\ket{\tilde{\psi}_{-,\downarrow}}|^2-|\bra{\tilde{\psi}_{-,\downarrow}}U(T_0)\ket{\tilde{\psi}_{-,\uparrow}^{(1)}}|^2, 
\end{align}
where $U_0(t)$ is the time-evolution operator for the effective Hamiltonian in Eq.~\eqref{Eq:SiliconEffH}. 
Moreover, for definiteness we will choose a particular realization of the $\pi$-pulse as follows, 
\begin{align}
\Omega(t) = \dfrac{w\pi}{4}\sin\dfrac{wt}{\hbar}. \label{Eq:Pi pulse}
\end{align}

\begin{figure}[!]
\includegraphics[scale=1.0]{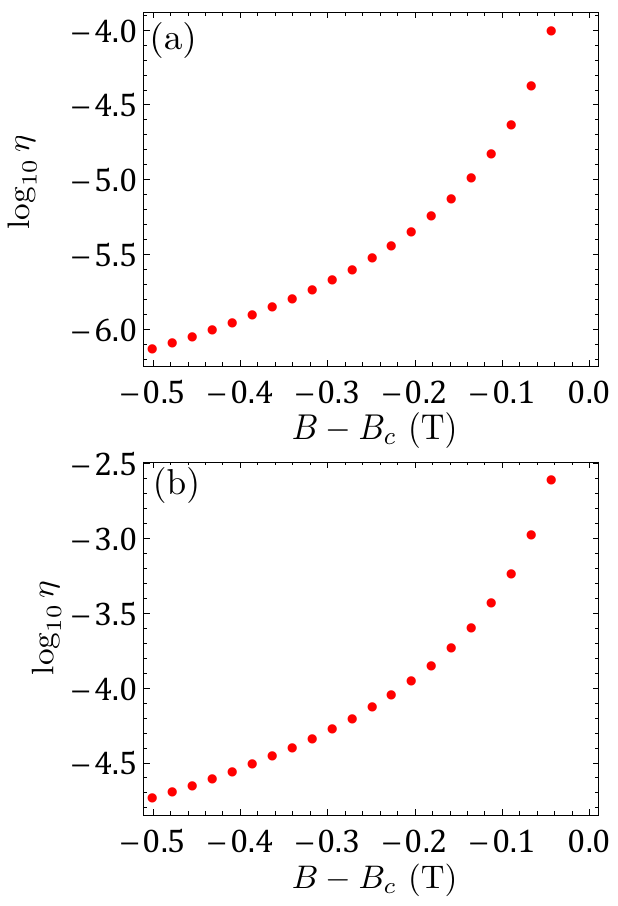}
\caption{\label{Fig:Silicon-Leakage-1} Leakage to the other valley eigenstate when the SOC strength $\Delta_a$ is (a) $\SI{0.1}{\micro eV}$ and (b) $\SI{0.5}{\micro eV}$, respectively. The detuning of the pulse is $\delta=\SI{0.1}{meV}$, and the valley splitting is $\EVS=\SI{0.33}{meV}$. The pulse duration is set to $T_0 = \pi\hbar/10\delta \simeq \SI{2.0}{ps}$. }
\end{figure}

Our numerical results find that the dominant factor that affects the leakage is $\xi$ [see Eq.~\eqref{Eq:RenormalizedStates}], the hybridization between $\tilde{\psi}_{-,\uparrow}$ and $\tilde{\psi}_{+,\downarrow}$, while the pulse parameter $\delta$ and duration $T_0$ have little impact, as long as the pulse is in the ``fast limit''. 
As shown in Fig.~\ref{Fig:Silicon-Leakage-1}, the leakage varies with $\xi$ through its dependence on the spin-orbit coupling strength and detuning from the hot spot $B_c$. 
In Fig.~\ref{Fig:Silicon-Leakage-1}(a) we plot the leakage using the SOC strength extracted from the recent experiment, $\Delta_a =\SI{0.1}{\micro eV}$~\cite{Yang2013:SiliconCubit}, and we find that the leakage quickly decreases below $10^{-4}$ when the applied field is $\SI{0.1}{T}$ away from $B_c$. In contrast, as shown in Fig.~\ref{Fig:Silicon-Leakage-1}(b), if we use an SOC strength five times larger, the same level of leakage can be achieved only when the field is about $\SI{0.25}{T}$ away from $B_c$. 
Meanwhile, given that the ability to tune valley splitting by a perpendicular electric field has been well demonstrated~\cite{Yang2013:SiliconCubit}, it is also possible to avoid such leakage hot spots by tuning the valley splittings. 
Thus, the leakage due to the valley-orbit and spin-orbit couplings can suppressed by tuning either an applied magnetic or electric field. 

The upshot of these results is that the leakage to the other valley subspace can be minimized if one avoids the leakage hot spots in the spin qubit operations; the exact amount of necessary detuning for obtaining a given low leakage level however, depends crucially on the strength of the spin-orbit coupling in the silicon quantum dot. 
Therefore we conclude that the leakage due to multiple valley states is usually not a serious concern for spin qubit operations in silicon, where spin-orbit and valley-orbit couplings are both small. 

\subsection{Discussions}
The leakage in this single-dot-double-valley model in Si  makes a nice comparison with that in the double-dot-single-valley model in GaAs. 
Indeed, the Hamiltonian for these two systems have a similar structure~[cf. Eq.~\eqref{Eq:SiliconEffH} for Si and Eq.~\eqref{Eq:WannierFourLevel} for GaAs]. Here we would like to contrast these two situations, which will help us gain some additional insights. 

The Hilbert space of these two problems can both be decomposed into a spin subspace and an orbital subspace. In the silicon problem, the orbital degrees of freedom arise from the two valleys in its band structure, which are completely decoupled from each other. As a result, in a pristine silicon quantum dot no leakage between the two valley states is possible, as long as there is a finite splitting between them. 
This scenario contrasts sharply with the case in GaAs, where the orbital degrees of freedom arise from the two Fock-Darwin states associated with the two quantum dots. In such a case a finite coupling always exists between the two FD states unless the two dots are infinitely far away from each other (i.e., in the $X_0/a \to \infty$ limit). This is the fundamental reason for leakage errors that occur in a GaAs double quantum dot. 

One may take a step further and ask why the two valleys in the silicon case are completely decoupled while the two FD states in the GaAs case are coupled? The reason is that, the two valley states are the eigenstates of their parent Hamiltonian, while the two FD states are not, as we now explain. 
If we compare the valley-orbit coupling term in Eq.~\eqref{Eq:ValleyOrbit} with the double quantum dot Hamiltonian at zero orbital detuning ($\tilde{\varepsilon}_d=0$) in Eq.~\eqref{Eq:WannierH}, it is clear that both Hamiltonians are proportional to $\sigma_x$. 
We further recall that the valley eigenstates are the exact eigenstates of $\sigma_x$, and thus they are completely decoupled. In contrast, the two FD states are not eigenstates of $\sigma_x$ (thus not eigenstates of their parent Hamiltonian), which explains why finite couplings between the two FD states still remain. 
If we were to define the spin qubits in terms of an equal superposition of the two FD states (an eigenstate of the double quantum dot Hamiltonian), the orbital dynamics of the spin qubit will always remain in this state, and no leakage would be possible, which is exactly what happens between the two valleys in silicon. 
Obviously, any leakage in a quantum problem is a statement about the distribution of wave function amplitude in different candidate states, and thus the magnitude of the leakage depends crucially on how these candidate states are defined.

We mention two additional aspects of the leakage problem in quantum dots here.  First, obviously all the features of double-dot spin transport leakage apply equally well to Si double quantum dot systems too, but Si has the additional element of multivalley physics which leads to a potential leakage even in the single-dot situation.  Second, in addition to the types of leakage considered in this work for quantum dot spin qubits, there could always be also leakage associated with the higher orbital confinement levels in each dot since quantum dots typically have many confined orbital levels.  This type of leakage associated with higher orbital levels (taking the system away from the pure two-level qubit subspace) is usually very small since the higher orbital levels are typically several meV above the ground qubit states~\cite{hu2002gate}.

\section{Summary \label{Section:Summary}}

In this work we study the leakage in transporting a single-spin qubit through a double quantum dot (and also, the complementary problem of leakage in a single quantum dot due to two valley states). We mainly focused on three types of leakage errors: finite ramping times, spin-dependent tunneling rates due to spin-orbit couplings, and the presence of multiple valley states. 
In the first and last case, we demonstrate that the leakage errors can be corrected or avoided relatively easily. Specifically, the ramping errors can be compensated by adjusting the tunneling time appropriately, while the leakage due to multiple valley states can be avoided by operating the spin qubit away from the ``leakage hot spots'', which occurs when the Zeeman splitting equals the valley splitting. 
For the second case where the leakage arises from the spin-dependent tunneling rates, we find that pulse shaping is generally needed to reduce the leakage errors. We discuss several possible optimal pulse shapes to minimize the leakage error. For example, in order to preserve the spin state during the transport, a carefully designed $2\pi$-pulse should be applied. 
To find such pulses we develop a formalism that can systematically analyze the leakage errors and generate the necessary pulse shapes to reduce the errors. Importantly, we show that universal, leakage-suppressing $2\pi$-pulses can be found without precisely knowing the spin state being transported. 
We further apply the same formalism to analyze leakage errors when certain spin rotations are performed during spin shuttling, and show that similar pulse shaping conditions can still be found, although they will generally depend on the initial spin state as well as the spin rotations performed.

\section{Acknowledgment}
This work is supported by LPS-MPO-CMTC. 

\bibliography{Bib-SpinQubit}

\begin{thebibliography}{58}%
\makeatletter
\providecommand \@ifxundefined [1]{%
 \@ifx{#1\undefined}
}%
\providecommand \@ifnum [1]{%
 \ifnum #1\expandafter \@firstoftwo
 \else \expandafter \@secondoftwo
 \fi
}%
\providecommand \@ifx [1]{%
 \ifx #1\expandafter \@firstoftwo
 \else \expandafter \@secondoftwo
 \fi
}%
\providecommand \natexlab [1]{#1}%
\providecommand \enquote  [1]{``#1''}%
\providecommand \bibnamefont  [1]{#1}%
\providecommand \bibfnamefont [1]{#1}%
\providecommand \citenamefont [1]{#1}%
\providecommand \href@noop [0]{\@secondoftwo}%
\providecommand \href [0]{\begingroup \@sanitize@url \@href}%
\providecommand \@href[1]{\@@startlink{#1}\@@href}%
\providecommand \@@href[1]{\endgroup#1\@@endlink}%
\providecommand \@sanitize@url [0]{\catcode `\\12\catcode `\$12\catcode
  `\&12\catcode `\#12\catcode `\^12\catcode `\_12\catcode `\%12\relax}%
\providecommand \@@startlink[1]{}%
\providecommand \@@endlink[0]{}%
\providecommand \url  [0]{\begingroup\@sanitize@url \@url }%
\providecommand \@url [1]{\endgroup\@href {#1}{\urlprefix }}%
\providecommand \urlprefix  [0]{URL }%
\providecommand \Eprint [0]{\href }%
\providecommand \doibase [0]{http://dx.doi.org/}%
\providecommand \selectlanguage [0]{\@gobble}%
\providecommand \bibinfo  [0]{\@secondoftwo}%
\providecommand \bibfield  [0]{\@secondoftwo}%
\providecommand \translation [1]{[#1]}%
\providecommand \BibitemOpen [0]{}%
\providecommand \bibitemStop [0]{}%
\providecommand \bibitemNoStop [0]{.\EOS\space}%
\providecommand \EOS [0]{\spacefactor3000\relax}%
\providecommand \BibitemShut  [1]{\csname bibitem#1\endcsname}%
\let\auto@bib@innerbib\@empty
\bibitem [{\citenamefont {Ono}\ \emph {et~al.}(2005)\citenamefont {Ono},
  \citenamefont {Fujiwara}, \citenamefont {Nishiguchi}, \citenamefont
  {Inokawa},\ and\ \citenamefont {Takahashi}}]{Ono2005}%
  \BibitemOpen
  \bibfield  {author} {\bibinfo {author} {\bibfnamefont {Y.}~\bibnamefont
  {Ono}}, \bibinfo {author} {\bibfnamefont {A.}~\bibnamefont {Fujiwara}},
  \bibinfo {author} {\bibfnamefont {K.}~\bibnamefont {Nishiguchi}}, \bibinfo
  {author} {\bibfnamefont {H.}~\bibnamefont {Inokawa}}, \ and\ \bibinfo
  {author} {\bibfnamefont {Y.}~\bibnamefont {Takahashi}},\ }\href {\doibase
  10.1063/1.1843271} {\bibfield  {journal} {\bibinfo  {journal} {J. Appl.
  Phys.}\ }\textbf {\bibinfo {volume} {97}},\ \bibinfo {pages} {031101}
  (\bibinfo {year} {2005})}\BibitemShut {NoStop}%
\bibitem [{\citenamefont {Kane}(1998)}]{Kane1998}%
  \BibitemOpen
  \bibfield  {author} {\bibinfo {author} {\bibfnamefont {B.~E.}\ \bibnamefont
  {Kane}},\ }\href {\doibase 10.1038/30156} {\bibfield  {journal} {\bibinfo
  {journal} {Nature}\ }\textbf {\bibinfo {volume} {393}},\ \bibinfo {pages}
  {133} (\bibinfo {year} {1998})}\BibitemShut {NoStop}%
\bibitem [{\citenamefont {Vrijen}\ \emph {et~al.}(2000)\citenamefont {Vrijen},
  \citenamefont {Yablonovitch}, \citenamefont {Wang}, \citenamefont {Jiang},
  \citenamefont {Balandin}, \citenamefont {Roychowdhury}, \citenamefont {Mor},\
  and\ \citenamefont {DiVincenzo}}]{Vrijen2000electron}%
  \BibitemOpen
  \bibfield  {author} {\bibinfo {author} {\bibfnamefont {R.}~\bibnamefont
  {Vrijen}}, \bibinfo {author} {\bibfnamefont {E.}~\bibnamefont
  {Yablonovitch}}, \bibinfo {author} {\bibfnamefont {K.}~\bibnamefont {Wang}},
  \bibinfo {author} {\bibfnamefont {H.~W.}\ \bibnamefont {Jiang}}, \bibinfo
  {author} {\bibfnamefont {A.}~\bibnamefont {Balandin}}, \bibinfo {author}
  {\bibfnamefont {V.}~\bibnamefont {Roychowdhury}}, \bibinfo {author}
  {\bibfnamefont {T.}~\bibnamefont {Mor}}, \ and\ \bibinfo {author}
  {\bibfnamefont {D.}~\bibnamefont {DiVincenzo}},\ }\href {\doibase
  10.1103/PhysRevA.62.012306} {\bibfield  {journal} {\bibinfo  {journal} {Phys.
  Rev. A}\ }\textbf {\bibinfo {volume} {62}},\ \bibinfo {pages} {012306}
  (\bibinfo {year} {2000})}\BibitemShut {NoStop}%
\bibitem [{\citenamefont {Loss}\ and\ \citenamefont
  {DiVincenzo}(1998)}]{Loss1998}%
  \BibitemOpen
  \bibfield  {author} {\bibinfo {author} {\bibfnamefont {D.}~\bibnamefont
  {Loss}}\ and\ \bibinfo {author} {\bibfnamefont {D.~P.}\ \bibnamefont
  {DiVincenzo}},\ }\href {\doibase 10.1103/PhysRevA.57.120} {\bibfield
  {journal} {\bibinfo  {journal} {Phys. Rev. A}\ }\textbf {\bibinfo {volume}
  {57}},\ \bibinfo {pages} {120} (\bibinfo {year} {1998})}\BibitemShut
  {NoStop}%
\bibitem [{\citenamefont {Oosterkamp}\ \emph {et~al.}(1998)\citenamefont
  {Oosterkamp}, \citenamefont {Fujisawa}, \citenamefont {van~der Wiel},
  \citenamefont {Ishibashi}, \citenamefont {Hijman}, \citenamefont {Tarucha},\
  and\ \citenamefont {Kouwenhoven}}]{Oosterkamp1998:QuantumDot}%
  \BibitemOpen
  \bibfield  {author} {\bibinfo {author} {\bibfnamefont {T.~H.}\ \bibnamefont
  {Oosterkamp}}, \bibinfo {author} {\bibfnamefont {T.}~\bibnamefont
  {Fujisawa}}, \bibinfo {author} {\bibfnamefont {W.~G.}\ \bibnamefont {van~der
  Wiel}}, \bibinfo {author} {\bibfnamefont {K.}~\bibnamefont {Ishibashi}},
  \bibinfo {author} {\bibfnamefont {R.~V.}\ \bibnamefont {Hijman}}, \bibinfo
  {author} {\bibfnamefont {S.}~\bibnamefont {Tarucha}}, \ and\ \bibinfo
  {author} {\bibfnamefont {L.~P.}\ \bibnamefont {Kouwenhoven}},\ }\href
  {\doibase 10.1038/27617} {\bibfield  {journal} {\bibinfo  {journal} {Nature}\
  }\textbf {\bibinfo {volume} {395}},\ \bibinfo {pages} {873} (\bibinfo {year}
  {1998})}\BibitemShut {NoStop}%
\bibitem [{\citenamefont {Burkard}\ \emph {et~al.}(1999)\citenamefont
  {Burkard}, \citenamefont {Loss},\ and\ \citenamefont
  {DiVincenzo}}]{Burkard1999:QuantumDots}%
  \BibitemOpen
  \bibfield  {author} {\bibinfo {author} {\bibfnamefont {G.}~\bibnamefont
  {Burkard}}, \bibinfo {author} {\bibfnamefont {D.}~\bibnamefont {Loss}}, \
  and\ \bibinfo {author} {\bibfnamefont {D.~P.}\ \bibnamefont {DiVincenzo}},\
  }\href {\doibase 10.1103/PhysRevB.59.2070} {\bibfield  {journal} {\bibinfo
  {journal} {Phys. Rev. B}\ }\textbf {\bibinfo {volume} {59}},\ \bibinfo
  {pages} {2070} (\bibinfo {year} {1999})}\BibitemShut {NoStop}%
\bibitem [{\citenamefont {Hu}\ and\ \citenamefont {{Das
  Sarma}}(2000)}]{Hu2000hilbert}%
  \BibitemOpen
  \bibfield  {author} {\bibinfo {author} {\bibfnamefont {X.}~\bibnamefont
  {Hu}}\ and\ \bibinfo {author} {\bibfnamefont {S.}~\bibnamefont {{Das
  Sarma}}},\ }\href {\doibase 10.1103/PhysRevA.61.062301} {\bibfield  {journal}
  {\bibinfo  {journal} {Phys. Rev. A}\ }\textbf {\bibinfo {volume} {61}},\
  \bibinfo {pages} {062301} (\bibinfo {year} {2000})}\BibitemShut {NoStop}%
\bibitem [{\citenamefont {Hu}\ and\ \citenamefont {{Das
  Sarma}}(2001)}]{Hu2001spin}%
  \BibitemOpen
  \bibfield  {author} {\bibinfo {author} {\bibfnamefont {X.}~\bibnamefont
  {Hu}}\ and\ \bibinfo {author} {\bibfnamefont {S.}~\bibnamefont {{Das
  Sarma}}},\ }\href {\doibase 10.1103/PhysRevA.64.042312} {\bibfield  {journal}
  {\bibinfo  {journal} {Phys. Rev. A}\ }\textbf {\bibinfo {volume} {64}},\
  \bibinfo {pages} {042312} (\bibinfo {year} {2001})}\BibitemShut {NoStop}%
\bibitem [{\citenamefont {Hu}\ and\ \citenamefont {{Das
  Sarma}}(2006)}]{Hu2006:SpinQubit}%
  \BibitemOpen
  \bibfield  {author} {\bibinfo {author} {\bibfnamefont {X.}~\bibnamefont
  {Hu}}\ and\ \bibinfo {author} {\bibfnamefont {S.}~\bibnamefont {{Das
  Sarma}}},\ }\href {\doibase 10.1103/PhysRevLett.96.100501} {\bibfield
  {journal} {\bibinfo  {journal} {Phys. Rev. Lett.}\ }\textbf {\bibinfo
  {volume} {96}},\ \bibinfo {pages} {100501} (\bibinfo {year}
  {2006})}\BibitemShut {NoStop}%
\bibitem [{\citenamefont {Shulman}\ \emph {et~al.}(2012)\citenamefont
  {Shulman}, \citenamefont {Dial}, \citenamefont {Harvey}, \citenamefont
  {Bluhm}, \citenamefont {Umansky},\ and\ \citenamefont
  {Yacoby}}]{Shulman2012}%
  \BibitemOpen
  \bibfield  {author} {\bibinfo {author} {\bibfnamefont {M.~D.}\ \bibnamefont
  {Shulman}}, \bibinfo {author} {\bibfnamefont {O.~E.}\ \bibnamefont {Dial}},
  \bibinfo {author} {\bibfnamefont {S.~P.}\ \bibnamefont {Harvey}}, \bibinfo
  {author} {\bibfnamefont {H.}~\bibnamefont {Bluhm}}, \bibinfo {author}
  {\bibfnamefont {V.}~\bibnamefont {Umansky}}, \ and\ \bibinfo {author}
  {\bibfnamefont {A.}~\bibnamefont {Yacoby}},\ }\href {\doibase
  10.1126/science.1217692} {\bibfield  {journal} {\bibinfo  {journal}
  {Science}\ }\textbf {\bibinfo {volume} {336}},\ \bibinfo {pages} {202}
  (\bibinfo {year} {2012})}\BibitemShut {NoStop}%
\bibitem [{\citenamefont {Bluhm}\ \emph {et~al.}(2011)\citenamefont {Bluhm},
  \citenamefont {Foletti}, \citenamefont {Neder}, \citenamefont {Rudner},
  \citenamefont {Mahalu}, \citenamefont {Umansky},\ and\ \citenamefont
  {Yacoby}}]{Bluhm2011}%
  \BibitemOpen
  \bibfield  {author} {\bibinfo {author} {\bibfnamefont {H.}~\bibnamefont
  {Bluhm}}, \bibinfo {author} {\bibfnamefont {S.}~\bibnamefont {Foletti}},
  \bibinfo {author} {\bibfnamefont {I.}~\bibnamefont {Neder}}, \bibinfo
  {author} {\bibfnamefont {M.}~\bibnamefont {Rudner}}, \bibinfo {author}
  {\bibfnamefont {D.}~\bibnamefont {Mahalu}}, \bibinfo {author} {\bibfnamefont
  {V.}~\bibnamefont {Umansky}}, \ and\ \bibinfo {author} {\bibfnamefont
  {A.}~\bibnamefont {Yacoby}},\ }\href {\doibase 10.1038/nphys1856} {\bibfield
  {journal} {\bibinfo  {journal} {Nat. Phys.}\ }\textbf {\bibinfo {volume}
  {7}},\ \bibinfo {pages} {109} (\bibinfo {year} {2011})}\BibitemShut {NoStop}%
\bibitem [{\citenamefont {Veldhorst}\ \emph {et~al.}(2014)\citenamefont
  {Veldhorst}, \citenamefont {Hwang}, \citenamefont {Yang}, \citenamefont
  {Leenstra}, \citenamefont {de~Ronde}, \citenamefont {Dehollain},
  \citenamefont {Muhonen}, \citenamefont {Hudson}, \citenamefont {Itoh},
  \citenamefont {Morello},\ and\ \citenamefont {Dzurak}}]{Veldhorst2014}%
  \BibitemOpen
  \bibfield  {author} {\bibinfo {author} {\bibfnamefont {M.}~\bibnamefont
  {Veldhorst}}, \bibinfo {author} {\bibfnamefont {J.~C.~C.}\ \bibnamefont
  {Hwang}}, \bibinfo {author} {\bibfnamefont {C.~H.}\ \bibnamefont {Yang}},
  \bibinfo {author} {\bibfnamefont {a.~W.}\ \bibnamefont {Leenstra}}, \bibinfo
  {author} {\bibfnamefont {B.}~\bibnamefont {de~Ronde}}, \bibinfo {author}
  {\bibfnamefont {J.~P.}\ \bibnamefont {Dehollain}}, \bibinfo {author}
  {\bibfnamefont {J.~T.}\ \bibnamefont {Muhonen}}, \bibinfo {author}
  {\bibfnamefont {F.~E.}\ \bibnamefont {Hudson}}, \bibinfo {author}
  {\bibfnamefont {K.~M.}\ \bibnamefont {Itoh}}, \bibinfo {author}
  {\bibfnamefont {A.}~\bibnamefont {Morello}}, \ and\ \bibinfo {author}
  {\bibfnamefont {a.~S.}\ \bibnamefont {Dzurak}},\ }\href {\doibase
  10.1038/nnano.2014.216} {\bibfield  {journal} {\bibinfo  {journal} {Nat.
  Nanotechnol.}\ }\textbf {\bibinfo {volume} {9}},\ \bibinfo {pages} {981}
  (\bibinfo {year} {2014})}\BibitemShut {NoStop}%
\bibitem [{\citenamefont {Braakman}\ \emph {et~al.}(2013)\citenamefont
  {Braakman}, \citenamefont {Barthelemy}, \citenamefont {Reichl}, \citenamefont
  {Wegscheider},\ and\ \citenamefont {Vandersypen}}]{Braakman2013}%
  \BibitemOpen
  \bibfield  {author} {\bibinfo {author} {\bibfnamefont {F.~R.}\ \bibnamefont
  {Braakman}}, \bibinfo {author} {\bibfnamefont {P.}~\bibnamefont
  {Barthelemy}}, \bibinfo {author} {\bibfnamefont {C.}~\bibnamefont {Reichl}},
  \bibinfo {author} {\bibfnamefont {W.}~\bibnamefont {Wegscheider}}, \ and\
  \bibinfo {author} {\bibfnamefont {L.}~\bibnamefont {Vandersypen}},\ }\href
  {\doibase 10.1038/nnano.2013.67} {\bibfield  {journal} {\bibinfo  {journal}
  {Nat. Nanotech.}\ }\textbf {\bibinfo {volume} {8}},\ \bibinfo {pages} {432}
  (\bibinfo {year} {2013})}\BibitemShut {NoStop}%
\bibitem [{\citenamefont {Takakura}\ \emph {et~al.}(2014)\citenamefont
  {Takakura}, \citenamefont {Noiri}, \citenamefont {Obata}, \citenamefont
  {Otsuka}, \citenamefont {Yoneda}, \citenamefont {Yoshida},\ and\
  \citenamefont {Tarucha}}]{Takakura2014}%
  \BibitemOpen
  \bibfield  {author} {\bibinfo {author} {\bibfnamefont {T.}~\bibnamefont
  {Takakura}}, \bibinfo {author} {\bibfnamefont {A.}~\bibnamefont {Noiri}},
  \bibinfo {author} {\bibfnamefont {T.}~\bibnamefont {Obata}}, \bibinfo
  {author} {\bibfnamefont {T.}~\bibnamefont {Otsuka}}, \bibinfo {author}
  {\bibfnamefont {J.}~\bibnamefont {Yoneda}}, \bibinfo {author} {\bibfnamefont
  {K.}~\bibnamefont {Yoshida}}, \ and\ \bibinfo {author} {\bibfnamefont
  {S.}~\bibnamefont {Tarucha}},\ }\href {\doibase 10.1063/1.4869108} {\bibfield
   {journal} {\bibinfo  {journal} {Appl. Phys. Lett.}\ }\textbf {\bibinfo
  {volume} {104}},\ \bibinfo {pages} {113109} (\bibinfo {year}
  {2014})}\BibitemShut {NoStop}%
\bibitem [{\citenamefont {DiVincenzo}\ \emph {et~al.}(2000)\citenamefont
  {DiVincenzo}, \citenamefont {Bacon}, \citenamefont {Kempe}, \citenamefont
  {Burkard},\ and\ \citenamefont {Whaley}}]{DiVincenzo2000}%
  \BibitemOpen
  \bibfield  {author} {\bibinfo {author} {\bibfnamefont {D.~P.}\ \bibnamefont
  {DiVincenzo}}, \bibinfo {author} {\bibfnamefont {D.}~\bibnamefont {Bacon}},
  \bibinfo {author} {\bibfnamefont {J.}~\bibnamefont {Kempe}}, \bibinfo
  {author} {\bibfnamefont {G.}~\bibnamefont {Burkard}}, \ and\ \bibinfo
  {author} {\bibfnamefont {K.~B.}\ \bibnamefont {Whaley}},\ }\href {\doibase
  10.1038/35042541} {\bibfield  {journal} {\bibinfo  {journal} {Nature}\
  }\textbf {\bibinfo {volume} {408}},\ \bibinfo {pages} {339} (\bibinfo {year}
  {2000})}\BibitemShut {NoStop}%
\bibitem [{\citenamefont {Hollenberg}\ \emph {et~al.}(2006)\citenamefont
  {Hollenberg}, \citenamefont {Greentree}, \citenamefont {Fowler},\ and\
  \citenamefont {Wellard}}]{Hollenberg2006}%
  \BibitemOpen
  \bibfield  {author} {\bibinfo {author} {\bibfnamefont {L.}~\bibnamefont
  {Hollenberg}}, \bibinfo {author} {\bibfnamefont {A.}~\bibnamefont
  {Greentree}}, \bibinfo {author} {\bibfnamefont {A.}~\bibnamefont {Fowler}}, \
  and\ \bibinfo {author} {\bibfnamefont {C.}~\bibnamefont {Wellard}},\ }\href
  {\doibase 10.1103/PhysRevB.74.045311} {\bibfield  {journal} {\bibinfo
  {journal} {Phys. Rev. B}\ }\textbf {\bibinfo {volume} {74}},\ \bibinfo
  {pages} {045311} (\bibinfo {year} {2006})}\BibitemShut {NoStop}%
\bibitem [{\citenamefont {Cole}\ \emph {et~al.}(2008)\citenamefont {Cole},
  \citenamefont {Greentree}, \citenamefont {Hollenberg},\ and\ \citenamefont
  {{Das Sarma}}}]{cole2008spatial}%
  \BibitemOpen
  \bibfield  {author} {\bibinfo {author} {\bibfnamefont {J.~H.}\ \bibnamefont
  {Cole}}, \bibinfo {author} {\bibfnamefont {A.~D.}\ \bibnamefont {Greentree}},
  \bibinfo {author} {\bibfnamefont {L.}~\bibnamefont {Hollenberg}}, \ and\
  \bibinfo {author} {\bibfnamefont {S.}~\bibnamefont {{Das Sarma}}},\ }\href
  {\doibase 10.1103/PhysRevB.77.235418} {\bibfield  {journal} {\bibinfo
  {journal} {Phys. Rev. B}\ }\textbf {\bibinfo {volume} {77}},\ \bibinfo
  {pages} {235418} (\bibinfo {year} {2008})}\BibitemShut {NoStop}%
\bibitem [{\citenamefont {Hill}\ \emph {et~al.}(2015)\citenamefont {Hill},
  \citenamefont {Peretz}, \citenamefont {Hile}, \citenamefont {House},
  \citenamefont {Fuechsle}, \citenamefont {Rogge}, \citenamefont {Simmons},\
  and\ \citenamefont {Hollenberg}}]{Hill2015}%
  \BibitemOpen
  \bibfield  {author} {\bibinfo {author} {\bibfnamefont {C.~D.}\ \bibnamefont
  {Hill}}, \bibinfo {author} {\bibfnamefont {E.}~\bibnamefont {Peretz}},
  \bibinfo {author} {\bibfnamefont {S.~J.}\ \bibnamefont {Hile}}, \bibinfo
  {author} {\bibfnamefont {M.~G.}\ \bibnamefont {House}}, \bibinfo {author}
  {\bibfnamefont {M.}~\bibnamefont {Fuechsle}}, \bibinfo {author}
  {\bibfnamefont {S.}~\bibnamefont {Rogge}}, \bibinfo {author} {\bibfnamefont
  {M.~Y.}\ \bibnamefont {Simmons}}, \ and\ \bibinfo {author} {\bibfnamefont
  {L.~C.}\ \bibnamefont {Hollenberg}},\ }\href {\doibase
  10.1126/sciadv.1500707} {\bibfield  {journal} {\bibinfo  {journal} {Sci.
  Adv.}\ }\textbf {\bibinfo {volume} {1}},\ \bibinfo {pages} {e1500707}
  (\bibinfo {year} {2015})}\BibitemShut {NoStop}%
\bibitem [{\citenamefont {McNeil}\ \emph {et~al.}(2011)\citenamefont {McNeil},
  \citenamefont {Kataoka}, \citenamefont {Ford}, \citenamefont {Barnes},
  \citenamefont {Anderson}, \citenamefont {Jones}, \citenamefont {Farrer},\
  and\ \citenamefont {Ritchie}}]{Mcneil2011}%
  \BibitemOpen
  \bibfield  {author} {\bibinfo {author} {\bibfnamefont {R.}~\bibnamefont
  {McNeil}}, \bibinfo {author} {\bibfnamefont {M.}~\bibnamefont {Kataoka}},
  \bibinfo {author} {\bibfnamefont {C.}~\bibnamefont {Ford}}, \bibinfo {author}
  {\bibfnamefont {C.}~\bibnamefont {Barnes}}, \bibinfo {author} {\bibfnamefont
  {D.}~\bibnamefont {Anderson}}, \bibinfo {author} {\bibfnamefont
  {G.}~\bibnamefont {Jones}}, \bibinfo {author} {\bibfnamefont
  {I.}~\bibnamefont {Farrer}}, \ and\ \bibinfo {author} {\bibfnamefont
  {D.}~\bibnamefont {Ritchie}},\ }\href {\doibase 10.1038/nature10444}
  {\bibfield  {journal} {\bibinfo  {journal} {Nature}\ }\textbf {\bibinfo
  {volume} {477}},\ \bibinfo {pages} {439} (\bibinfo {year}
  {2011})}\BibitemShut {NoStop}%
\bibitem [{\citenamefont {Baart}\ \emph {et~al.}()\citenamefont {Baart},
  \citenamefont {Shafiei}, \citenamefont {Fujita}, \citenamefont {Reichl},
  \citenamefont {Wegscheider},\ and\ \citenamefont {Vandersypen}}]{Baart2016}%
  \BibitemOpen
  \bibfield  {author} {\bibinfo {author} {\bibfnamefont {T.~A.}\ \bibnamefont
  {Baart}}, \bibinfo {author} {\bibfnamefont {M.}~\bibnamefont {Shafiei}},
  \bibinfo {author} {\bibfnamefont {T.}~\bibnamefont {Fujita}}, \bibinfo
  {author} {\bibfnamefont {C.}~\bibnamefont {Reichl}}, \bibinfo {author}
  {\bibfnamefont {W.}~\bibnamefont {Wegscheider}}, \ and\ \bibinfo {author}
  {\bibfnamefont {L.}~\bibnamefont {Vandersypen}},\ }\href {\doibase
  10.1038/nnano.2015.291} {\bibfield  {journal} {\bibinfo  {journal} {Nat.
  Nanotech. {advance online publication}, 04 January (2016)}\
  }10.1038/nnano.2015.291}\BibitemShut {NoStop}%
\bibitem [{\citenamefont {Taylor}\ \emph {et~al.}(2005)\citenamefont {Taylor},
  \citenamefont {Engel}, \citenamefont {D{\"u}r}, \citenamefont {Yacoby},
  \citenamefont {Marcus}, \citenamefont {Zoller},\ and\ \citenamefont
  {Lukin}}]{Taylor2005}%
  \BibitemOpen
  \bibfield  {author} {\bibinfo {author} {\bibfnamefont {J.}~\bibnamefont
  {Taylor}}, \bibinfo {author} {\bibfnamefont {H.-A.}\ \bibnamefont {Engel}},
  \bibinfo {author} {\bibfnamefont {W.}~\bibnamefont {D{\"u}r}}, \bibinfo
  {author} {\bibfnamefont {A.}~\bibnamefont {Yacoby}}, \bibinfo {author}
  {\bibfnamefont {C.}~\bibnamefont {Marcus}}, \bibinfo {author} {\bibfnamefont
  {P.}~\bibnamefont {Zoller}}, \ and\ \bibinfo {author} {\bibfnamefont
  {M.}~\bibnamefont {Lukin}},\ }\href {\doibase 10.1038/nphys174} {\bibfield
  {journal} {\bibinfo  {journal} {Nat. Phys.}\ }\textbf {\bibinfo {volume}
  {1}},\ \bibinfo {pages} {177} (\bibinfo {year} {2005})}\BibitemShut {NoStop}%
\bibitem [{\citenamefont {Fowler}\ \emph {et~al.}(2012)\citenamefont {Fowler},
  \citenamefont {Mariantoni}, \citenamefont {Martinis},\ and\ \citenamefont
  {Cleland}}]{Fowler2012}%
  \BibitemOpen
  \bibfield  {author} {\bibinfo {author} {\bibfnamefont {A.~G.}\ \bibnamefont
  {Fowler}}, \bibinfo {author} {\bibfnamefont {M.}~\bibnamefont {Mariantoni}},
  \bibinfo {author} {\bibfnamefont {J.~M.}\ \bibnamefont {Martinis}}, \ and\
  \bibinfo {author} {\bibfnamefont {A.~N.}\ \bibnamefont {Cleland}},\ }\href
  {\doibase 10.1103/PhysRevA.86.032324} {\bibfield  {journal} {\bibinfo
  {journal} {Phys. Rev. A}\ }\textbf {\bibinfo {volume} {86}},\ \bibinfo
  {pages} {032324} (\bibinfo {year} {2012})}\BibitemShut {NoStop}%
\bibitem [{\citenamefont {O'Gorman}\ \emph {et~al.}()\citenamefont {O'Gorman},
  \citenamefont {Nickerson}, \citenamefont {Ross}, \citenamefont {Morton},\
  and\ \citenamefont {Benjamin}}]{OGorman2014}%
  \BibitemOpen
  \bibfield  {author} {\bibinfo {author} {\bibfnamefont {J.}~\bibnamefont
  {O'Gorman}}, \bibinfo {author} {\bibfnamefont {N.~H.}\ \bibnamefont
  {Nickerson}}, \bibinfo {author} {\bibfnamefont {P.}~\bibnamefont {Ross}},
  \bibinfo {author} {\bibfnamefont {J.~J.}\ \bibnamefont {Morton}}, \ and\
  \bibinfo {author} {\bibfnamefont {S.~C.}\ \bibnamefont {Benjamin}},\
  }\href@noop {} {\ }\Eprint {http://arxiv.org/abs/1406.5149} {arXiv:1406.5149}
  \BibitemShut {NoStop}%
\bibitem [{\citenamefont {Pica}\ \emph {et~al.}()\citenamefont {Pica},
  \citenamefont {Lovett}, \citenamefont {Bhatt}, \citenamefont {Schenkel},\
  and\ \citenamefont {Lyon}}]{Pica2015}%
  \BibitemOpen
  \bibfield  {author} {\bibinfo {author} {\bibfnamefont {G.}~\bibnamefont
  {Pica}}, \bibinfo {author} {\bibfnamefont {B.}~\bibnamefont {Lovett}},
  \bibinfo {author} {\bibfnamefont {R.}~\bibnamefont {Bhatt}}, \bibinfo
  {author} {\bibfnamefont {T.}~\bibnamefont {Schenkel}}, \ and\ \bibinfo
  {author} {\bibfnamefont {S.}~\bibnamefont {Lyon}},\ }\href@noop {} {\
  }\Eprint {http://arxiv.org/abs/1506.04913} {arXiv:1506.04913} \BibitemShut
  {NoStop}%
\bibitem [{\citenamefont {Bulaev}\ and\ \citenamefont
  {Loss}(2005)}]{Bulaev2005:QuantumDot}%
  \BibitemOpen
  \bibfield  {author} {\bibinfo {author} {\bibfnamefont {D.~V.}\ \bibnamefont
  {Bulaev}}\ and\ \bibinfo {author} {\bibfnamefont {D.}~\bibnamefont {Loss}},\
  }\href {\doibase 10.1103/PhysRevB.71.205324} {\bibfield  {journal} {\bibinfo
  {journal} {Phys. Rev. B}\ }\textbf {\bibinfo {volume} {71}},\ \bibinfo
  {pages} {205324} (\bibinfo {year} {2005})}\BibitemShut {NoStop}%
\bibitem [{\citenamefont {Stano}\ and\ \citenamefont
  {Fabian}(2005)}]{Stano2005:QuantumDot}%
  \BibitemOpen
  \bibfield  {author} {\bibinfo {author} {\bibfnamefont {P.}~\bibnamefont
  {Stano}}\ and\ \bibinfo {author} {\bibfnamefont {J.}~\bibnamefont {Fabian}},\
  }\href {\doibase 10.1103/PhysRevB.72.155410} {\bibfield  {journal} {\bibinfo
  {journal} {Phys. Rev. B}\ }\textbf {\bibinfo {volume} {72}},\ \bibinfo
  {pages} {155410} (\bibinfo {year} {2005})}\BibitemShut {NoStop}%
\bibitem [{\citenamefont {Danon}\ and\ \citenamefont
  {Nazarov}(2009)}]{Danon2009}%
  \BibitemOpen
  \bibfield  {author} {\bibinfo {author} {\bibfnamefont {J.}~\bibnamefont
  {Danon}}\ and\ \bibinfo {author} {\bibfnamefont {Y.~V.}\ \bibnamefont
  {Nazarov}},\ }\href {\doibase 10.1103/PhysRevB.80.041301} {\bibfield
  {journal} {\bibinfo  {journal} {Phys. Rev. B}\ }\textbf {\bibinfo {volume}
  {80}},\ \bibinfo {pages} {041301} (\bibinfo {year} {2009})}\BibitemShut
  {NoStop}%
\bibitem [{\citenamefont {Schreiber}\ \emph {et~al.}(2011)\citenamefont
  {Schreiber}, \citenamefont {Braakman}, \citenamefont {Meunier}, \citenamefont
  {Calado}, \citenamefont {Danon}, \citenamefont {Taylor}, \citenamefont
  {Wegscheider},\ and\ \citenamefont {Vandersypen}}]{Schreiber2011}%
  \BibitemOpen
  \bibfield  {author} {\bibinfo {author} {\bibfnamefont {L.}~\bibnamefont
  {Schreiber}}, \bibinfo {author} {\bibfnamefont {F.}~\bibnamefont {Braakman}},
  \bibinfo {author} {\bibfnamefont {T.}~\bibnamefont {Meunier}}, \bibinfo
  {author} {\bibfnamefont {V.}~\bibnamefont {Calado}}, \bibinfo {author}
  {\bibfnamefont {J.}~\bibnamefont {Danon}}, \bibinfo {author} {\bibfnamefont
  {J.}~\bibnamefont {Taylor}}, \bibinfo {author} {\bibfnamefont
  {W.}~\bibnamefont {Wegscheider}}, \ and\ \bibinfo {author} {\bibfnamefont
  {L.}~\bibnamefont {Vandersypen}},\ }\href {\doibase 10.1038/ncomms1561}
  {\bibfield  {journal} {\bibinfo  {journal} {Nat. Commun.}\ }\textbf {\bibinfo
  {volume} {2}},\ \bibinfo {pages} {556} (\bibinfo {year} {2011})}\BibitemShut
  {NoStop}%
\bibitem [{\citenamefont {Stepanenko}\ \emph {et~al.}(2012)\citenamefont
  {Stepanenko}, \citenamefont {Rudner}, \citenamefont {Halperin},\ and\
  \citenamefont {Loss}}]{Stepanenko2012}%
  \BibitemOpen
  \bibfield  {author} {\bibinfo {author} {\bibfnamefont {D.}~\bibnamefont
  {Stepanenko}}, \bibinfo {author} {\bibfnamefont {M.}~\bibnamefont {Rudner}},
  \bibinfo {author} {\bibfnamefont {B.~I.}\ \bibnamefont {Halperin}}, \ and\
  \bibinfo {author} {\bibfnamefont {D.}~\bibnamefont {Loss}},\ }\href {\doibase
  10.1103/PhysRevB.85.075416} {\bibfield  {journal} {\bibinfo  {journal} {Phys.
  Rev. B}\ }\textbf {\bibinfo {volume} {85}},\ \bibinfo {pages} {075416}
  (\bibinfo {year} {2012})}\BibitemShut {NoStop}%
\bibitem [{\citenamefont {Yang}\ \emph {et~al.}(2013)\citenamefont {Yang},
  \citenamefont {Rossi}, \citenamefont {Ruskov}, \citenamefont {Lai},
  \citenamefont {Mohiyaddin}, \citenamefont {Lee}, \citenamefont {Tahan},
  \citenamefont {Klimeck}, \citenamefont {Morello},\ and\ \citenamefont
  {Dzurak}}]{Yang2013:SiliconCubit}%
  \BibitemOpen
  \bibfield  {author} {\bibinfo {author} {\bibfnamefont {C.~H.}\ \bibnamefont
  {Yang}}, \bibinfo {author} {\bibfnamefont {A.}~\bibnamefont {Rossi}},
  \bibinfo {author} {\bibfnamefont {R.}~\bibnamefont {Ruskov}}, \bibinfo
  {author} {\bibfnamefont {N.~S.}\ \bibnamefont {Lai}}, \bibinfo {author}
  {\bibfnamefont {F.~A.}\ \bibnamefont {Mohiyaddin}}, \bibinfo {author}
  {\bibfnamefont {S.}~\bibnamefont {Lee}}, \bibinfo {author} {\bibfnamefont
  {C.}~\bibnamefont {Tahan}}, \bibinfo {author} {\bibfnamefont
  {G.}~\bibnamefont {Klimeck}}, \bibinfo {author} {\bibfnamefont
  {A.}~\bibnamefont {Morello}}, \ and\ \bibinfo {author} {\bibfnamefont
  {A.~S.}\ \bibnamefont {Dzurak}},\ }\href {\doibase 10.1038/ncomms3069}
  {\bibfield  {journal} {\bibinfo  {journal} {Nat. Commun.}\ }\textbf {\bibinfo
  {volume} {4}},\ \bibinfo {pages} {2069} (\bibinfo {year} {2013})}\BibitemShut
  {NoStop}%
\bibitem [{\citenamefont {Maisi}\ \emph {et~al.}()\citenamefont {Maisi},
  \citenamefont {Hofmann}, \citenamefont {Roosli}, \citenamefont {Basset},
  \citenamefont {Reichl}, \citenamefont {Wegscheider}, \citenamefont {Ihn},\
  and\ \citenamefont {Ensslin}}]{Maisi2015:SpinQubit}%
  \BibitemOpen
  \bibfield  {author} {\bibinfo {author} {\bibfnamefont {V.~F.}\ \bibnamefont
  {Maisi}}, \bibinfo {author} {\bibfnamefont {A.}~\bibnamefont {Hofmann}},
  \bibinfo {author} {\bibfnamefont {M.}~\bibnamefont {Roosli}}, \bibinfo
  {author} {\bibfnamefont {J.}~\bibnamefont {Basset}}, \bibinfo {author}
  {\bibfnamefont {C.}~\bibnamefont {Reichl}}, \bibinfo {author} {\bibfnamefont
  {W.}~\bibnamefont {Wegscheider}}, \bibinfo {author} {\bibfnamefont
  {T.}~\bibnamefont {Ihn}}, \ and\ \bibinfo {author} {\bibfnamefont
  {K.}~\bibnamefont {Ensslin}},\ }\href {http://arxiv.org/abs/1512.05149} {\
  }\Eprint {http://arxiv.org/abs/1512.05149} {arXiv:1512.05149} \BibitemShut
  {NoStop}%
\bibitem [{\citenamefont {Veldhorst}\ \emph
  {et~al.}(2015{\natexlab{a}})\citenamefont {Veldhorst}, \citenamefont
  {Ruskov}, \citenamefont {Yang}, \citenamefont {Hwang}, \citenamefont
  {Hudson}, \citenamefont {Flatt{\'e}}, \citenamefont {Tahan}, \citenamefont
  {Itoh}, \citenamefont {Morello},\ and\ \citenamefont
  {Dzurak}}]{Veldhorst2015SOC}%
  \BibitemOpen
  \bibfield  {author} {\bibinfo {author} {\bibfnamefont {M.}~\bibnamefont
  {Veldhorst}}, \bibinfo {author} {\bibfnamefont {R.}~\bibnamefont {Ruskov}},
  \bibinfo {author} {\bibfnamefont {C.}~\bibnamefont {Yang}}, \bibinfo {author}
  {\bibfnamefont {J.}~\bibnamefont {Hwang}}, \bibinfo {author} {\bibfnamefont
  {F.}~\bibnamefont {Hudson}}, \bibinfo {author} {\bibfnamefont
  {M.}~\bibnamefont {Flatt{\'e}}}, \bibinfo {author} {\bibfnamefont
  {C.}~\bibnamefont {Tahan}}, \bibinfo {author} {\bibfnamefont
  {K.}~\bibnamefont {Itoh}}, \bibinfo {author} {\bibfnamefont {A.}~\bibnamefont
  {Morello}}, \ and\ \bibinfo {author} {\bibfnamefont {A.}~\bibnamefont
  {Dzurak}},\ }\href {\doibase 10.1103/PhysRevB.92.201401} {\bibfield
  {journal} {\bibinfo  {journal} {Phys. Rev. B}\ }\textbf {\bibinfo {volume}
  {92}},\ \bibinfo {pages} {201401} (\bibinfo {year}
  {2015}{\natexlab{a}})}\BibitemShut {NoStop}%
\bibitem [{\citenamefont {Boykin}\ \emph {et~al.}(2004)\citenamefont {Boykin},
  \citenamefont {Klimeck}, \citenamefont {Friesen}, \citenamefont
  {Coppersmith}, \citenamefont {von Allmen}, \citenamefont {Oyafuso},\ and\
  \citenamefont {Lee}}]{Boykin2004:ValleySplitting}%
  \BibitemOpen
  \bibfield  {author} {\bibinfo {author} {\bibfnamefont {T.~B.}\ \bibnamefont
  {Boykin}}, \bibinfo {author} {\bibfnamefont {G.}~\bibnamefont {Klimeck}},
  \bibinfo {author} {\bibfnamefont {M.}~\bibnamefont {Friesen}}, \bibinfo
  {author} {\bibfnamefont {S.~N.}\ \bibnamefont {Coppersmith}}, \bibinfo
  {author} {\bibfnamefont {P.}~\bibnamefont {von Allmen}}, \bibinfo {author}
  {\bibfnamefont {F.}~\bibnamefont {Oyafuso}}, \ and\ \bibinfo {author}
  {\bibfnamefont {S.}~\bibnamefont {Lee}},\ }\href {\doibase
  10.1103/PhysRevB.70.165325} {\bibfield  {journal} {\bibinfo  {journal} {Phys.
  Rev. B}\ }\textbf {\bibinfo {volume} {70}},\ \bibinfo {pages} {165325}
  (\bibinfo {year} {2004})}\BibitemShut {NoStop}%
\bibitem [{\citenamefont {Culcer}\ \emph {et~al.}(2009)\citenamefont {Culcer},
  \citenamefont {Cywinski}, \citenamefont {Li}, \citenamefont {Hu},\ and\
  \citenamefont {{Das Sarma}}}]{Culcer2009:SpinQubit}%
  \BibitemOpen
  \bibfield  {author} {\bibinfo {author} {\bibfnamefont {D.}~\bibnamefont
  {Culcer}}, \bibinfo {author} {\bibfnamefont {L.}~\bibnamefont {Cywinski}},
  \bibinfo {author} {\bibfnamefont {Q.}~\bibnamefont {Li}}, \bibinfo {author}
  {\bibfnamefont {X.}~\bibnamefont {Hu}}, \ and\ \bibinfo {author}
  {\bibfnamefont {S.}~\bibnamefont {{Das Sarma}}},\ }\href {\doibase
  10.1103/PhysRevB.80.205302} {\bibfield  {journal} {\bibinfo  {journal} {Phys.
  Rev. B}\ }\textbf {\bibinfo {volume} {80}},\ \bibinfo {pages} {205302}
  (\bibinfo {year} {2009})}\BibitemShut {NoStop}%
\bibitem [{\citenamefont {Saraiva}\ \emph {et~al.}(2009)\citenamefont
  {Saraiva}, \citenamefont {Calderon}, \citenamefont {Hu}, \citenamefont {{Das
  Sarma}},\ and\ \citenamefont {Koiller}}]{Saraiva2009:SpinQubit}%
  \BibitemOpen
  \bibfield  {author} {\bibinfo {author} {\bibfnamefont {A.~L.}\ \bibnamefont
  {Saraiva}}, \bibinfo {author} {\bibfnamefont {M.~J.}\ \bibnamefont
  {Calderon}}, \bibinfo {author} {\bibfnamefont {X.}~\bibnamefont {Hu}},
  \bibinfo {author} {\bibfnamefont {S.}~\bibnamefont {{Das Sarma}}}, \ and\
  \bibinfo {author} {\bibfnamefont {B.}~\bibnamefont {Koiller}},\ }\href
  {\doibase 10.1103/PhysRevB.80.081305} {\bibfield  {journal} {\bibinfo
  {journal} {Phys. Rev. B}\ }\textbf {\bibinfo {volume} {80}},\ \bibinfo
  {pages} {081305} (\bibinfo {year} {2009})}\BibitemShut {NoStop}%
\bibitem [{\citenamefont {Li}\ \emph {et~al.}(2010)\citenamefont {Li},
  \citenamefont {Cywinski}, \citenamefont {Culcer}, \citenamefont {Hu},\ and\
  \citenamefont {{Das Sarma}}}]{Li2010:Qubits}%
  \BibitemOpen
  \bibfield  {author} {\bibinfo {author} {\bibfnamefont {Q.}~\bibnamefont
  {Li}}, \bibinfo {author} {\bibfnamefont {L.}~\bibnamefont {Cywinski}},
  \bibinfo {author} {\bibfnamefont {D.}~\bibnamefont {Culcer}}, \bibinfo
  {author} {\bibfnamefont {X.}~\bibnamefont {Hu}}, \ and\ \bibinfo {author}
  {\bibfnamefont {S.}~\bibnamefont {{Das Sarma}}},\ }\href {\doibase
  10.1103/PhysRevB.81.085313} {\bibfield  {journal} {\bibinfo  {journal} {Phys.
  Rev. B}\ }\textbf {\bibinfo {volume} {81}},\ \bibinfo {pages} {085313}
  (\bibinfo {year} {2010})}\BibitemShut {NoStop}%
\bibitem [{\citenamefont {Culcer}\ \emph
  {et~al.}(2010{\natexlab{a}})\citenamefont {Culcer}, \citenamefont {Hu},\ and\
  \citenamefont {{Das Sarma}}}]{Culcer2010:SiliconQubit}%
  \BibitemOpen
  \bibfield  {author} {\bibinfo {author} {\bibfnamefont {D.}~\bibnamefont
  {Culcer}}, \bibinfo {author} {\bibfnamefont {X.}~\bibnamefont {Hu}}, \ and\
  \bibinfo {author} {\bibfnamefont {S.}~\bibnamefont {{Das Sarma}}},\ }\href
  {\doibase 10.1103/PhysRevB.82.205315} {\bibfield  {journal} {\bibinfo
  {journal} {Phys. Rev. B}\ }\textbf {\bibinfo {volume} {82}},\ \bibinfo
  {pages} {205315} (\bibinfo {year} {2010}{\natexlab{a}})}\BibitemShut
  {NoStop}%
\bibitem [{\citenamefont {Friesen}\ and\ \citenamefont
  {Coppersmith}(2010)}]{Friesen2010:valley-orbit}%
  \BibitemOpen
  \bibfield  {author} {\bibinfo {author} {\bibfnamefont {M.}~\bibnamefont
  {Friesen}}\ and\ \bibinfo {author} {\bibfnamefont {S.~N.}\ \bibnamefont
  {Coppersmith}},\ }\href {\doibase 10.1103/PhysRevB.81.115324} {\bibfield
  {journal} {\bibinfo  {journal} {Phys. Rev. B}\ }\textbf {\bibinfo {volume}
  {81}},\ \bibinfo {pages} {115324} (\bibinfo {year} {2010})}\BibitemShut
  {NoStop}%
\bibitem [{\citenamefont {Culcer}\ \emph
  {et~al.}(2010{\natexlab{b}})\citenamefont {Culcer}, \citenamefont {Cywinski},
  \citenamefont {Li}, \citenamefont {Hu},\ and\ \citenamefont {{Das
  Sarma}}}]{Culcer2010:Cubit}%
  \BibitemOpen
  \bibfield  {author} {\bibinfo {author} {\bibfnamefont {D.}~\bibnamefont
  {Culcer}}, \bibinfo {author} {\bibfnamefont {L.}~\bibnamefont {Cywinski}},
  \bibinfo {author} {\bibfnamefont {Q.}~\bibnamefont {Li}}, \bibinfo {author}
  {\bibfnamefont {X.}~\bibnamefont {Hu}}, \ and\ \bibinfo {author}
  {\bibfnamefont {S.}~\bibnamefont {{Das Sarma}}},\ }\href {\doibase
  10.1103/PhysRevB.82.155312} {\bibfield  {journal} {\bibinfo  {journal} {Phys.
  Rev. B}\ }\textbf {\bibinfo {volume} {82}},\ \bibinfo {pages} {155312}
  (\bibinfo {year} {2010}{\natexlab{b}})}\BibitemShut {NoStop}%
\bibitem [{\citenamefont {Saraiva}\ \emph {et~al.}(2011)\citenamefont
  {Saraiva}, \citenamefont {Calderon}, \citenamefont {Capaz}, \citenamefont
  {Hu}, \citenamefont {{Das Sarma}},\ and\ \citenamefont
  {Koiller}}]{Saraiva2011:SpinQubit}%
  \BibitemOpen
  \bibfield  {author} {\bibinfo {author} {\bibfnamefont {A.~L.}\ \bibnamefont
  {Saraiva}}, \bibinfo {author} {\bibfnamefont {M.~J.}\ \bibnamefont
  {Calderon}}, \bibinfo {author} {\bibfnamefont {R.~B.}\ \bibnamefont {Capaz}},
  \bibinfo {author} {\bibfnamefont {X.}~\bibnamefont {Hu}}, \bibinfo {author}
  {\bibfnamefont {S.}~\bibnamefont {{Das Sarma}}}, \ and\ \bibinfo {author}
  {\bibfnamefont {B.}~\bibnamefont {Koiller}},\ }\href {\doibase
  10.1103/PhysRevB.84.155320} {\bibfield  {journal} {\bibinfo  {journal} {Phys.
  Rev. B}\ }\textbf {\bibinfo {volume} {84}},\ \bibinfo {pages} {155320}
  (\bibinfo {year} {2011})}\BibitemShut {NoStop}%
\bibitem [{\citenamefont {Tyryshkin}\ \emph {et~al.}(2006)\citenamefont
  {Tyryshkin}, \citenamefont {Morton}, \citenamefont {Benjamin}, \citenamefont
  {Ardavan}, \citenamefont {Briggs}, \citenamefont {Ager},\ and\ \citenamefont
  {Lyon}}]{Tyryshkin2006}%
  \BibitemOpen
  \bibfield  {author} {\bibinfo {author} {\bibfnamefont {A.}~\bibnamefont
  {Tyryshkin}}, \bibinfo {author} {\bibfnamefont {J.}~\bibnamefont {Morton}},
  \bibinfo {author} {\bibfnamefont {S.}~\bibnamefont {Benjamin}}, \bibinfo
  {author} {\bibfnamefont {A.}~\bibnamefont {Ardavan}}, \bibinfo {author}
  {\bibfnamefont {G.}~\bibnamefont {Briggs}}, \bibinfo {author} {\bibfnamefont
  {J.}~\bibnamefont {Ager}}, \ and\ \bibinfo {author} {\bibfnamefont
  {S.}~\bibnamefont {Lyon}},\ }\href {\doibase 10.1088/0953-8984/18/21/S06}
  {\bibfield  {journal} {\bibinfo  {journal} {J. Phys. Condens. Matter}\
  }\textbf {\bibinfo {volume} {18}},\ \bibinfo {pages} {S783} (\bibinfo {year}
  {2006})}\BibitemShut {NoStop}%
\bibitem [{\citenamefont {Pla}\ \emph {et~al.}(2012)\citenamefont {Pla},
  \citenamefont {Tan}, \citenamefont {Dehollain}, \citenamefont {Lim},
  \citenamefont {Morton}, \citenamefont {Jamieson}, \citenamefont {Dzurak},\
  and\ \citenamefont {Morello}}]{Pla2012}%
  \BibitemOpen
  \bibfield  {author} {\bibinfo {author} {\bibfnamefont {J.~J.}\ \bibnamefont
  {Pla}}, \bibinfo {author} {\bibfnamefont {K.~Y.}\ \bibnamefont {Tan}},
  \bibinfo {author} {\bibfnamefont {J.~P.}\ \bibnamefont {Dehollain}}, \bibinfo
  {author} {\bibfnamefont {W.~H.}\ \bibnamefont {Lim}}, \bibinfo {author}
  {\bibfnamefont {J.~J.}\ \bibnamefont {Morton}}, \bibinfo {author}
  {\bibfnamefont {D.~N.}\ \bibnamefont {Jamieson}}, \bibinfo {author}
  {\bibfnamefont {A.~S.}\ \bibnamefont {Dzurak}}, \ and\ \bibinfo {author}
  {\bibfnamefont {A.}~\bibnamefont {Morello}},\ }\href {\doibase
  10.1038/nature11449} {\bibfield  {journal} {\bibinfo  {journal} {Nature}\
  }\textbf {\bibinfo {volume} {489}},\ \bibinfo {pages} {541} (\bibinfo {year}
  {2012})}\BibitemShut {NoStop}%
\bibitem [{\citenamefont {Kawakami}\ \emph {et~al.}(2014)\citenamefont
  {Kawakami}, \citenamefont {Scarlino}, \citenamefont {Ward}, \citenamefont
  {Braakman}, \citenamefont {Savage}, \citenamefont {Lagally}, \citenamefont
  {Friesen}, \citenamefont {Coppersmith}, \citenamefont {Eriksson},\ and\
  \citenamefont {Vandersypen}}]{Kawakami2014}%
  \BibitemOpen
  \bibfield  {author} {\bibinfo {author} {\bibfnamefont {E.}~\bibnamefont
  {Kawakami}}, \bibinfo {author} {\bibfnamefont {P.}~\bibnamefont {Scarlino}},
  \bibinfo {author} {\bibfnamefont {D.}~\bibnamefont {Ward}}, \bibinfo {author}
  {\bibfnamefont {F.}~\bibnamefont {Braakman}}, \bibinfo {author}
  {\bibfnamefont {D.}~\bibnamefont {Savage}}, \bibinfo {author} {\bibfnamefont
  {M.}~\bibnamefont {Lagally}}, \bibinfo {author} {\bibfnamefont
  {M.}~\bibnamefont {Friesen}}, \bibinfo {author} {\bibfnamefont
  {S.}~\bibnamefont {Coppersmith}}, \bibinfo {author} {\bibfnamefont
  {M.}~\bibnamefont {Eriksson}}, \ and\ \bibinfo {author} {\bibfnamefont
  {L.}~\bibnamefont {Vandersypen}},\ }\href {\doibase 10.1038/nnano.2014.153}
  {\bibfield  {journal} {\bibinfo  {journal} {Nat. Nanotechnol.}\ }\textbf
  {\bibinfo {volume} {9}},\ \bibinfo {pages} {666} (\bibinfo {year}
  {2014})}\BibitemShut {NoStop}%
\bibitem [{\citenamefont {Veldhorst}\ \emph
  {et~al.}(2015{\natexlab{b}})\citenamefont {Veldhorst}, \citenamefont {Yang},
  \citenamefont {Hwang}, \citenamefont {Huang}, \citenamefont {Dehollain},
  \citenamefont {Muhonen}, \citenamefont {Simmons}, \citenamefont {Laucht},
  \citenamefont {Hudson}, \citenamefont {Itoh}, \citenamefont {Morello},\ and\
  \citenamefont {Dzurak}}]{Veldhorst2015}%
  \BibitemOpen
  \bibfield  {author} {\bibinfo {author} {\bibfnamefont {M.}~\bibnamefont
  {Veldhorst}}, \bibinfo {author} {\bibfnamefont {C.}~\bibnamefont {Yang}},
  \bibinfo {author} {\bibfnamefont {J.}~\bibnamefont {Hwang}}, \bibinfo
  {author} {\bibfnamefont {W.}~\bibnamefont {Huang}}, \bibinfo {author}
  {\bibfnamefont {J.}~\bibnamefont {Dehollain}}, \bibinfo {author}
  {\bibfnamefont {J.}~\bibnamefont {Muhonen}}, \bibinfo {author} {\bibfnamefont
  {S.}~\bibnamefont {Simmons}}, \bibinfo {author} {\bibfnamefont
  {A.}~\bibnamefont {Laucht}}, \bibinfo {author} {\bibfnamefont
  {F.}~\bibnamefont {Hudson}}, \bibinfo {author} {\bibfnamefont
  {K.}~\bibnamefont {Itoh}}, \bibinfo {author} {\bibfnamefont {A.}~\bibnamefont
  {Morello}}, \ and\ \bibinfo {author} {\bibfnamefont {A.~S.}\ \bibnamefont
  {Dzurak}},\ }\href {\doibase 10.1038/nature15263} {\bibfield  {journal}
  {\bibinfo  {journal} {Nature}\ }\textbf {\bibinfo {volume} {526}},\ \bibinfo
  {pages} {410} (\bibinfo {year} {2015}{\natexlab{b}})}\BibitemShut {NoStop}%
\bibitem [{\citenamefont {Zwanenburg}\ \emph {et~al.}(2013)\citenamefont
  {Zwanenburg}, \citenamefont {Dzurak}, \citenamefont {Morello}, \citenamefont
  {Simmons}, \citenamefont {Hollenberg}, \citenamefont {Klimeck}, \citenamefont
  {Rogge}, \citenamefont {Coppersmith},\ and\ \citenamefont
  {Eriksson}}]{Zwanenburg2013silicon}%
  \BibitemOpen
  \bibfield  {author} {\bibinfo {author} {\bibfnamefont {F.~A.}\ \bibnamefont
  {Zwanenburg}}, \bibinfo {author} {\bibfnamefont {A.~S.}\ \bibnamefont
  {Dzurak}}, \bibinfo {author} {\bibfnamefont {A.}~\bibnamefont {Morello}},
  \bibinfo {author} {\bibfnamefont {M.~Y.}\ \bibnamefont {Simmons}}, \bibinfo
  {author} {\bibfnamefont {L.~C.}\ \bibnamefont {Hollenberg}}, \bibinfo
  {author} {\bibfnamefont {G.}~\bibnamefont {Klimeck}}, \bibinfo {author}
  {\bibfnamefont {S.}~\bibnamefont {Rogge}}, \bibinfo {author} {\bibfnamefont
  {S.~N.}\ \bibnamefont {Coppersmith}}, \ and\ \bibinfo {author} {\bibfnamefont
  {M.~A.}\ \bibnamefont {Eriksson}},\ }\href {\doibase
  10.1103/RevModPhys.85.961} {\bibfield  {journal} {\bibinfo  {journal} {Rev.
  Mod. Phys.}\ }\textbf {\bibinfo {volume} {85}},\ \bibinfo {pages} {961}
  (\bibinfo {year} {2013})}\BibitemShut {NoStop}%
\bibitem [{\citenamefont {Itoh}\ and\ \citenamefont
  {Watanabe}(2014)}]{Itoh2014isotope}%
  \BibitemOpen
  \bibfield  {author} {\bibinfo {author} {\bibfnamefont {K.~M.}\ \bibnamefont
  {Itoh}}\ and\ \bibinfo {author} {\bibfnamefont {H.}~\bibnamefont
  {Watanabe}},\ }\href {\doibase 10.1557/mrc.2014.32} {\bibfield  {journal}
  {\bibinfo  {journal} {MRS Commun.}\ }\textbf {\bibinfo {volume} {4}},\
  \bibinfo {pages} {143} (\bibinfo {year} {2014})}\BibitemShut {NoStop}%
\bibitem [{\citenamefont {Maune}\ \emph {et~al.}(2012)\citenamefont {Maune},
  \citenamefont {Borselli}, \citenamefont {Huang}, \citenamefont {Ladd},
  \citenamefont {Deelman}, \citenamefont {Holabird}, \citenamefont {Kiselev},
  \citenamefont {Alvarado-Rodriguez}, \citenamefont {Ross}, \citenamefont
  {Schmitz} \emph {et~al.}}]{Maune2012}%
  \BibitemOpen
  \bibfield  {author} {\bibinfo {author} {\bibfnamefont {B.}~\bibnamefont
  {Maune}}, \bibinfo {author} {\bibfnamefont {M.}~\bibnamefont {Borselli}},
  \bibinfo {author} {\bibfnamefont {B.}~\bibnamefont {Huang}}, \bibinfo
  {author} {\bibfnamefont {T.}~\bibnamefont {Ladd}}, \bibinfo {author}
  {\bibfnamefont {P.}~\bibnamefont {Deelman}}, \bibinfo {author} {\bibfnamefont
  {K.}~\bibnamefont {Holabird}}, \bibinfo {author} {\bibfnamefont
  {A.}~\bibnamefont {Kiselev}}, \bibinfo {author} {\bibfnamefont
  {I.}~\bibnamefont {Alvarado-Rodriguez}}, \bibinfo {author} {\bibfnamefont
  {R.}~\bibnamefont {Ross}}, \bibinfo {author} {\bibfnamefont {A.}~\bibnamefont
  {Schmitz}},  \emph {et~al.},\ }\href {\doibase 10.1038/nature10707}
  {\bibfield  {journal} {\bibinfo  {journal} {Nature}\ }\textbf {\bibinfo
  {volume} {481}},\ \bibinfo {pages} {344} (\bibinfo {year}
  {2012})}\BibitemShut {NoStop}%
\bibitem [{\citenamefont {Kim}\ \emph {et~al.}(2014)\citenamefont {Kim},
  \citenamefont {Shi}, \citenamefont {Simmons}, \citenamefont {Ward},
  \citenamefont {Prance}, \citenamefont {Koh}, \citenamefont {Gamble},
  \citenamefont {Savage}, \citenamefont {Lagally}, \citenamefont {Friesen}
  \emph {et~al.}}]{Kim2014}%
  \BibitemOpen
  \bibfield  {author} {\bibinfo {author} {\bibfnamefont {D.}~\bibnamefont
  {Kim}}, \bibinfo {author} {\bibfnamefont {Z.}~\bibnamefont {Shi}}, \bibinfo
  {author} {\bibfnamefont {C.}~\bibnamefont {Simmons}}, \bibinfo {author}
  {\bibfnamefont {D.}~\bibnamefont {Ward}}, \bibinfo {author} {\bibfnamefont
  {J.}~\bibnamefont {Prance}}, \bibinfo {author} {\bibfnamefont {T.~S.}\
  \bibnamefont {Koh}}, \bibinfo {author} {\bibfnamefont {J.~K.}\ \bibnamefont
  {Gamble}}, \bibinfo {author} {\bibfnamefont {D.}~\bibnamefont {Savage}},
  \bibinfo {author} {\bibfnamefont {M.}~\bibnamefont {Lagally}}, \bibinfo
  {author} {\bibfnamefont {M.}~\bibnamefont {Friesen}},  \emph {et~al.},\
  }\href {\doibase 10.1038/nature13407} {\bibfield  {journal} {\bibinfo
  {journal} {Nature}\ }\textbf {\bibinfo {volume} {511}},\ \bibinfo {pages}
  {70} (\bibinfo {year} {2014})}\BibitemShut {NoStop}%
\bibitem [{\citenamefont {Bruus}\ and\ \citenamefont
  {Flensberg}(2004)}]{bruus2004many}%
  \BibitemOpen
  \bibfield  {author} {\bibinfo {author} {\bibfnamefont {H.}~\bibnamefont
  {Bruus}}\ and\ \bibinfo {author} {\bibfnamefont {K.}~\bibnamefont
  {Flensberg}},\ }\href@noop {} {\emph {\bibinfo {title} {Many-Body Quantum
  Theory in Condensed Matter Physics: An Introduction}}},\ Oxford Graduate
  Texts\ (\bibinfo  {publisher} {OUP Oxford},\ \bibinfo {year}
  {2004})\BibitemShut {NoStop}%
\bibitem [{\citenamefont {Barnes}\ and\ \citenamefont {{Das
  Sarma}}(2012)}]{Barnes2012:Analytical}%
  \BibitemOpen
  \bibfield  {author} {\bibinfo {author} {\bibfnamefont {E.}~\bibnamefont
  {Barnes}}\ and\ \bibinfo {author} {\bibfnamefont {S.}~\bibnamefont {{Das
  Sarma}}},\ }\href {\doibase 10.1103/PhysRevLett.109.060401} {\bibfield
  {journal} {\bibinfo  {journal} {Phys. Rev. Lett.}\ }\textbf {\bibinfo
  {volume} {109}},\ \bibinfo {pages} {060401} (\bibinfo {year}
  {2012})}\BibitemShut {NoStop}%
\bibitem [{\citenamefont {Barnes}\ \emph {et~al.}(2015)\citenamefont {Barnes},
  \citenamefont {Wang},\ and\ \citenamefont {{Das
  Sarma}}}]{Barnes2015:SpinCubit}%
  \BibitemOpen
  \bibfield  {author} {\bibinfo {author} {\bibfnamefont {E.}~\bibnamefont
  {Barnes}}, \bibinfo {author} {\bibfnamefont {X.}~\bibnamefont {Wang}}, \ and\
  \bibinfo {author} {\bibfnamefont {S.}~\bibnamefont {{Das Sarma}}},\ }\href
  {\doibase 10.1038/srep12685} {\bibfield  {journal} {\bibinfo  {journal} {Sci.
  Rep.}\ }\textbf {\bibinfo {volume} {5}},\ \bibinfo {pages} {12685} (\bibinfo
  {year} {2015})}\BibitemShut {NoStop}%
\bibitem [{\citenamefont {Li}\ \emph {et~al.}(2013)\citenamefont {Li},
  \citenamefont {Zhang},\ and\ \citenamefont {Niu}}]{LiXiao2013:MoS2}%
  \BibitemOpen
  \bibfield  {author} {\bibinfo {author} {\bibfnamefont {X.}~\bibnamefont
  {Li}}, \bibinfo {author} {\bibfnamefont {F.}~\bibnamefont {Zhang}}, \ and\
  \bibinfo {author} {\bibfnamefont {Q.}~\bibnamefont {Niu}},\ }\href {\doibase
  10.1103/PhysRevLett.110.066803} {\bibfield  {journal} {\bibinfo  {journal}
  {Phys. Rev. Lett.}\ }\textbf {\bibinfo {volume} {110}},\ \bibinfo {pages}
  {066803} (\bibinfo {year} {2013})}\BibitemShut {NoStop}%
\bibitem [{\citenamefont {Li}\ \emph {et~al.}(2014)\citenamefont {Li},
  \citenamefont {Zhang}, \citenamefont {Niu},\ and\ \citenamefont
  {MacDonald}}]{LiXiao2014:DomainWall}%
  \BibitemOpen
  \bibfield  {author} {\bibinfo {author} {\bibfnamefont {X.}~\bibnamefont
  {Li}}, \bibinfo {author} {\bibfnamefont {F.}~\bibnamefont {Zhang}}, \bibinfo
  {author} {\bibfnamefont {Q.}~\bibnamefont {Niu}}, \ and\ \bibinfo {author}
  {\bibfnamefont {A.~H.}\ \bibnamefont {MacDonald}},\ }\href {\doibase
  10.1103/PhysRevLett.113.116803} {\bibfield  {journal} {\bibinfo  {journal}
  {Phys. Rev. Lett.}\ }\textbf {\bibinfo {volume} {113}},\ \bibinfo {pages}
  {116803} (\bibinfo {year} {2014})}\BibitemShut {NoStop}%
\bibitem [{\citenamefont {Li}\ \emph {et~al.}(2015)\citenamefont {Li},
  \citenamefont {Roy},\ and\ \citenamefont {{Das Sarma}}}]{LiXiao2015:SmB6}%
  \BibitemOpen
  \bibfield  {author} {\bibinfo {author} {\bibfnamefont {X.}~\bibnamefont
  {Li}}, \bibinfo {author} {\bibfnamefont {B.}~\bibnamefont {Roy}}, \ and\
  \bibinfo {author} {\bibfnamefont {S.}~\bibnamefont {{Das Sarma}}},\ }\href
  {\doibase 10.1103/PhysRevB.92.235144} {\bibfield  {journal} {\bibinfo
  {journal} {Phys. Rev. B}\ }\textbf {\bibinfo {volume} {92}},\ \bibinfo
  {pages} {235144} (\bibinfo {year} {2015})}\BibitemShut {NoStop}%
\bibitem [{\citenamefont {Li}\ \emph {et~al.}(2016)\citenamefont {Li},
  \citenamefont {Zhang},\ and\ \citenamefont {MacDonald}}]{LiXiao2016:SnTe}%
  \BibitemOpen
  \bibfield  {author} {\bibinfo {author} {\bibfnamefont {X.}~\bibnamefont
  {Li}}, \bibinfo {author} {\bibfnamefont {F.}~\bibnamefont {Zhang}}, \ and\
  \bibinfo {author} {\bibfnamefont {A.~H.}\ \bibnamefont {MacDonald}},\ }\href
  {\doibase 10.1103/PhysRevLett.116.026803} {\bibfield  {journal} {\bibinfo
  {journal} {Phys. Rev. Lett.}\ }\textbf {\bibinfo {volume} {116}},\ \bibinfo
  {pages} {026803} (\bibinfo {year} {2016})}\BibitemShut {NoStop}%
\bibitem [{\citenamefont {{Das Sarma}}\ \emph {et~al.}(1979)\citenamefont {{Das
  Sarma}}, \citenamefont {Kalia}, \citenamefont {Nakayama},\ and\ \citenamefont
  {Quinn}}]{DasSarma1979:stress}%
  \BibitemOpen
  \bibfield  {author} {\bibinfo {author} {\bibfnamefont {S.}~\bibnamefont {{Das
  Sarma}}}, \bibinfo {author} {\bibfnamefont {R.~K.}\ \bibnamefont {Kalia}},
  \bibinfo {author} {\bibfnamefont {M.}~\bibnamefont {Nakayama}}, \ and\
  \bibinfo {author} {\bibfnamefont {J.~J.}\ \bibnamefont {Quinn}},\ }\href
  {\doibase 10.1103/PhysRevB.19.6397} {\bibfield  {journal} {\bibinfo
  {journal} {Phys. Rev. B}\ }\textbf {\bibinfo {volume} {19}},\ \bibinfo
  {pages} {6397} (\bibinfo {year} {1979})}\BibitemShut {NoStop}%
\bibitem [{\citenamefont {Stern}\ and\ \citenamefont {{Das
  Sarma}}(1984)}]{Stern1984electron}%
  \BibitemOpen
  \bibfield  {author} {\bibinfo {author} {\bibfnamefont {F.}~\bibnamefont
  {Stern}}\ and\ \bibinfo {author} {\bibfnamefont {S.}~\bibnamefont {{Das
  Sarma}}},\ }\href {\doibase 10.1103/PhysRevB.30.840} {\bibfield  {journal}
  {\bibinfo  {journal} {Phys. Rev. B}\ }\textbf {\bibinfo {volume} {30}},\
  \bibinfo {pages} {840} (\bibinfo {year} {1984})}\BibitemShut {NoStop}%
\bibitem [{\citenamefont {Hu}\ and\ \citenamefont {{Das
  Sarma}}(2002)}]{hu2002gate}%
  \BibitemOpen
  \bibfield  {author} {\bibinfo {author} {\bibfnamefont {X.}~\bibnamefont
  {Hu}}\ and\ \bibinfo {author} {\bibfnamefont {S.}~\bibnamefont {{Das
  Sarma}}},\ }\href {\doibase 10.1103/PhysRevA.66.012312} {\bibfield  {journal}
  {\bibinfo  {journal} {Phys. Rev. A}\ }\textbf {\bibinfo {volume} {66}},\
  \bibinfo {pages} {012312} (\bibinfo {year} {2002})}\BibitemShut {NoStop}%
\end{thebibliography}%
\end{document}